\documentclass[preprint,12pt]{elsarticle}

\usepackage{amssymb}
\usepackage{amsmath}
\usepackage[total={7.3in,10in}]{geometry}
\usepackage[separate-uncertainty=true, multi-part-units=single]{siunitx}
\usepackage{import}
\usepackage{graphicx}
\usepackage{sidecap}

\usepackage{easyReview}
\usepackage{hyperref}

\hypersetup{
	colorlinks=true,
	linkcolor= blue,
	filecolor=blue,      
	urlcolor=blue,
	citecolor=blue,
}
\AtBeginDocument{\hypersetup{citecolor=blue, linkcolor=blue, urlcolor=blue}}

\begin{document}

\begin{frontmatter}


\title{Entropic crystallization of Brownian squares through pathways governed by orientational dynamics}

\author[a]{Debojit Chanda}
\author[b,c]{Thomas G. Mason}
\author[a]{Manas Khan\corref{*}} 

\affiliation[a]{organization={Department of Physics, Indian Institute of Technology Kanpur}, city={Kanpur}, postcode={208016}, country={India}}
\affiliation[b]{organization={Department of Physics and Astronomy, University of California}, city={Los Angeles}, state={CA}, postcode={90095}, country={USA}}
\affiliation[c]{organization={Department of Chemistry and Biochemistry, University of California}, city={Los Angeles}, state={CA}, postcode={90095}, country={USA}}
\cortext[*]{mkhan@iitk.ac.in}


\begin{abstract}
In dense systems of hard-interacting colloidal particles having anisotropic shapes, crystallization pathways represent an interesting frontier. The translational and rotational dynamics of such particles become coupled at higher densities, resulting in complex kinetics of their configurational ordering. To elucidate this, we have studied a two-dimensional entropic system of osmotically compressed corner-rounded Brownian square platelets. By analyzing the translational and orientational dynamics of the particles and their respective contributions toward minimizing the free energy, we show that the range of accessible orientational states of the particles principally governs the pathways of structural evolution, as the orientational entropy dictates the minimization of the free energy and, hence, the resulting optimal equilibrium ordering. When the particles have access to a wider range of orientational states, the larger rotational component of configurational entropy minimizes the total free energy, leading to hexagonal ordering. At higher osmotic pressures, the long collective translational fluctuations of the side-aligned particles with restricted rotational fluctuations maximize the entropy with a greater contribution from the translational component, thereby inducing a free energetically favored rhombic crystalline structure. We further show that density influences the crystallization pathways indirectly by setting an upper bound on the range of accessible orientational states. Complementary Brownian dynamics simulations and free-energy calculations further corroborate our findings, and their generalizability is demonstrated using a system of triangular particles. Thus, orientational dynamics is predicted to play a crucial role in governing the pathways for entropic ordering of various anisotropic shapes.
\end{abstract}



\begin{keyword}
Crystallization pathways \sep Anisotropic colloids \sep Entropic crystallization \sep Colloidal crystals \sep  Square particles 



\end{keyword}

\end{frontmatter}

\section*{Introduction}

Kinetic pathways of crystallization are of universal importance because they dictate the structural evolution, and thus govern the characteristics of the emergent equilibrium phases in systems ranging from atomic to mesoscopic length scales \cite{DeYoreo2015,Loh2016,Li2016,Ou2019}. Hence, a comprehensive understanding of the pathways provides critical information on the conditions required for the onset and regulation of the crystallization process. These pathways play a crucial role in a wide range of applications, including material synthesis \cite{Whitesides2002,DeYoreo2015}, mineralization \cite{Henzler2018}, fabrication of optical metamaterials \cite{John1987,Stebe2009,Hensley2021}, organic solar cells  \cite{Xin2022}, protein crystallization \cite{McPherson1999,Dale2003}, pharmaceuticals \cite{Chen2011}, and electronics \cite{Choi2016}. Remarkably, in systems with directional short-range interactions such as patchy colloids \cite{Wilber2007,Yi2013}, anisotropic molecules \cite{DeYoreo2015}, nanoparticles \cite{Xia2011,Ye2016,Ou2019}, or hard-interacting micrometer-sized constituents \cite{Adams1998,WOJCIECHOWSKI2004,Valignat2005,Glotzer2007, Zhao2009,Stebe2009,Agarwal2011,Zhao2011, Damasceno2012,Avendano2012,Zhao2012,Zhao2012a,Wang2015,Liu2022}, kinetically and energetically preferred crystallization pathways often lead to a variety of equilibrium orderings, some of which are dramatically different. Extensive studies on the crystallization dynamics of similar systems have associated nonclassical multistep nucleation pathways with the formation of intermediate metastable phases and diverse equilibrium structures \cite{DeYoreo2015,Loh2016,Chen2018,Ou2019,Lee2019,Doan2024}.

Starting with Onsager’s spherocylinders  \cite{Onsager1949}, anisotropic hard-interacting colloids have been shown to equilibrate into a variety of structural orderings, including mesophases \cite{John2008,Agarwal2011,Shen2019}, liquid crystalline \cite{Onsager1949,Frenkel1988,MartinezRaton2005, John2008,Zhao2012a,Damasceno2012}, and crystalline phases \cite{WOJCIECHOWSKI2004,John2008,Zhao2009,Zhao2011,Damasceno2011,Zhao2012,Damasceno2012,Wang2015,Klotsa2018} with varying symmetries, depending on the geometry and density of the particles. However, the underlying crystallization dynamics remain largely unexplored because of the intricacy of these systems \cite{Lee2019,Doan2024}. It has been demonstrated that at higher densities, when the orientational excluded volume of the anisotropic constituents begins to overlap, the configurational entropy, which drives the structural ordering of hard-interacting particles, becomes dependent on shape symmetry and is called shape entropy \cite{Anders2013, Anders2014, Rocha2020, Lim2023}. This shape entropy is maximized when anisotropic shapes form facet-aligned structures induced by effective directional entropic forces (DEF) \cite{Damasceno2011,Damasceno2012, Anders2014,Harper2019,Rocha2020, Vo2022, Yang2023}. Hence, the underlying reorganization dynamics of anisotropic particles through their coupled translational-rotational motions and subsequent structural evolution govern the accessible routes to eventual equilibrium crystalline phases in such systems. Understanding these crystallization pathways is essential for discerning and engineering self-assemblies of hard-interacting anisotropic building blocks, such as proteins, viruses, custom-shaped colloidal particles, and other anisotropic constituents with targeted mechanical and photonic properties \cite{Adams1998,McPherson1999,Fry1999,Whitesides2002,Dale2003, Glotzer2007,Stebe2009,Agarwal2011, Damasceno2012}. Furthermore, the validity of the universality in density-governed crystallization in hard-sphere systems \cite{Alder1957,Wood1957,Hansen1991} remains to be tested experimentally for their anisotropic equivalents.

Dense colloidal assemblies of hard squares provide the most pertinent system for investigating the crystallization dynamics of anisotropic particles because of their intriguing equilibrium phase behavior, despite having one of the simplest polygon geometries. The exploration of the equilibrium phases of squares commenced with the calculation of higher-order virial coefficients \cite{Tarjus1991}. A Monte Carlo (MC) simulation study later predicted the melting of a square crystalline phase into a tetratic mesophase having fourfold orientational order as the density decreased \cite{WOJCIECHOWSKI2004}. However, experiments using a two-dimensional (2D) system of Brownian square platelets under osmotic compression showed strikingly different results. An entropic transition from the isotropic phase to the hexagonal rotator crystal (RX), and then to the rhombic crystalline phase (RB), with a coexistence phase (CE) in between, was observed and validated by a cage-like mean-field calculation \cite{Zhao2011}. It was argued that the slightly rounded corners of the lithographically printed squares caused them to stabilize into RB rather than square crystal at higher densities, reconciling the gap \cite{Zhao2011}. Later, it was corroborated with an MC study that showed the variation in equilibrium phases with varied corner rounding \cite{Avendano2012}. These studies strongly imply that the particles' configurational dynamics, which are greatly facilitated by the ability of corner crossing, regulate the pathways to achieve optimal crystalline ordering in equilibrium. Furthermore, the coexistence of neighboring small local RX and RB crystallites in the CE phase \cite{Zhao2011} poses an intriguing question: what drives the formation and stabilization of ordered structures with different symmetries?

Here, we studied the optimal reorganization dynamics of square particles, specifically, the relative contributions of translational and orientational motions therein, inducing structural evolution to maximize the configurational entropy and thus minimize the free energy in a densely packed assembly approaching crystallization. Using the same system as that used in a previous experimental study \cite{Zhao2011}, we tracked the translational and rotational dynamics of lithographic square platelets under osmotic compression, in 2D, where both dynamics are distinctly visible. Here, we focus our attention particularly on two regions: the boundaries of the coexistence phase with RB and RX, where the squares equilibrate to RB and the hexagonal crystalline phase (HX), respectively. The particles have restricted rotational motion in the HX, which is formed at a slightly higher density than in the RX, where the particles enjoy full orientational freedom. By comparing the translational and rotational dynamics of the squares in both the pre-crystalline and crystalline domains of RB and HX, and calculating their contributions to the change in free energy, we elucidated that the range of accessible rotational states of the particles governs the crystallization pathways, as the orientational component of the configurational entropy guides the minimization of the total free energy in this system, dictating whether they equilibrate into RB or HX at a given osmotic pressure. Our Brownian dynamics simulations on a relatively smaller size of the same system of squares, resembling the local structural domains observed in the experiments, exhibit the time evolution of their structural configurations, corroborating this conclusion.

It is important to note that the local density indeed regulates the upper limit of the range of accessible orientational fluctuations of particles when the rotational excluded areas start to overlap, and thereby can indirectly govern the equilibrium phase behavior, which is manifested in the crystal-crystal transition with varying area fractions in the same system of Brownian squares \cite{Zhao2011}. However, our results establish that if the range of accessible orientational states can be modulated independently of the local density with appropriate corner rounding of the Brownian squares, it solely dictates the route to crystallization and the consequent equilibrium ordering, akin to the density in hard-sphere systems. We further explain the applicability of our results to other dense systems of anisotropic hard-interacting particles, using triangular platelets \cite{Zhao2012a} as an explicit example.

\section*{Results}

\subsection*{Structural Ordering}

\begin{figure}[thb] 
	\centering
	\includegraphics[width=0.75\linewidth]{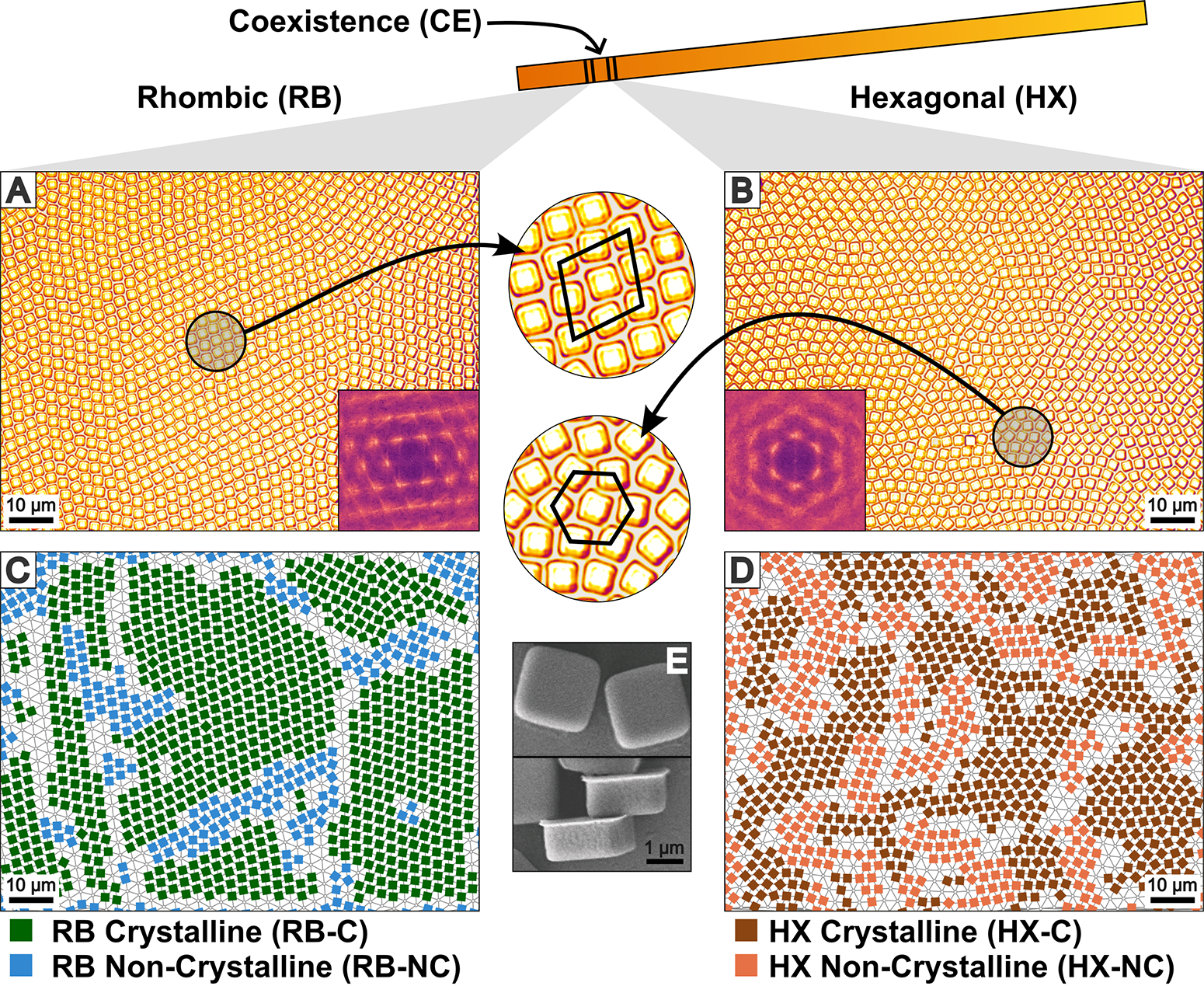}
	\caption{Dense 2D systems of hard-interacting square particles, forming rhombic (RB) and hexagonal (HX) phases, crystalline and non-crystalline domains therein. The top row shows bright-field optical images of the RB (A) and HX (B) phases formed at increasing heights, separated by a coexistence (CE) region, along an inclined capillary (not in scale), corresponding to decreasing osmotic pressure. Fourier transforms exhibiting the prevailing global ordering are shown in the insets. Magnified local regions of RB and HX are highlighted in the middle with symmetry structure (black polygons) overlays. Pseudo color is added to enhance clarity. The C and NC regions in the RB and HX phases, identified using local bond-orientational and positional order parameters (see text for identification criteria) of the time-averaged particle positions, are shown with the corresponding Delaunay triangulations in the bottom row (C and D, respectively). Particles forming C regions are denoted by green and brown squares, whereas those constituting NC regions are represented by blue and orange squares, for RB and HX, respectively. The square markers are rotated based on the mean particle orientation. Small grey circles mark the centers of particles that do not satisfy the criteria to be included in either the C or NC regions. Scanning electron micrographs show the face and side views of the square platelet particles (E).}
	\label{fig:System}
\end{figure}

Square platelet particles with marginally rounded corners formed a monolayer in 2D confinement on the flat bottom surface of an inclined rectangular cuvette and crystallized under gravity-induced osmotic compression along the length of the cuvette. These platelets lay flat on the bottom surface and can still move freely because of a carefully adjusted roughness controlled depletion attraction \cite{Zhao2007} between the particle face and glass surface. However, the in-plane depletion and van der Waals interactions are interaction is less than $k_{\text{B}}T$ in strength because of the rougher particle edges (roughness $\geq$ \SI{50}{nm}) and longer inter-particle separations ($\gtrsim$ \SI{300}{nm}), respectively, and are effectively overcome by thermal fluctuations. The hydrodynamic drag on lithographic particles is dominated by the lubrication interaction between the cuvette wall and the particle face nearer to it, resulting in reduced in-plane mobility. This slows down the dynamics and consequently the crystallization process, whereas the crystallization pathways remain solely dependent on the free-energy landscape, which, for our system of entropic colloids, depends only on the  configurational entropy of the system. Furthermore, the interparticle hydrodynamic interaction remains insignificant in influencing the crystallization dynamics and has therefore been neglected (SM). Near the bottom end, the dense system of squares equilibrated into two distinctly different symmetry structures, rhombic (RB) and hexagonal (HX), with decreasing osmotic pressure (Fig. \ref{fig:System}). Consequently, the particle density, defined by the area fraction ($\Phi$) in 2D, was marginally higher in RB ($\Phi_ {\text{RB}}$ = 0.782 $\pm$ 0.001) than in HX ($\Phi_ {\text{HX}}$ = 0.768 $\pm$ 0.001) as these two phases were separated by a narrow coexistence (CE) region. Because the configurational dynamics of the particles in this entropic system are very slow, progressing over several weeks, we studied the system at an intermediate time when the crystallization process was underway and both crystalline (C) and non-crystalline (NC) domains, which were yet to crystallize, coexisted. Both the translational and orientational dynamics of the squares in the RB and HX were tracked at 15 fps over a time interval of \SI{160}{s}, which was sufficiently large to provide adequate temporal statistics yet short enough to ensure the stationarity of the system during the observation period.

While RB and HX are distinguished qualitatively by the symmetry revealed by Fourier transforms (Fig. \ref{fig:System} (A) \& (B) Insets), global $m$-fold bond-orientational  ($\Psi_m$) and positional ($\text{Z}_m$) order parameters, calculated from the time-averaged center positions of the squares, provide quantitative structural distinctions (SM). RB is recognized by higher values of global 4-fold order parameters ($\Psi_4$ = 0.63 $\pm$ 0.01, $\text{Z}_4$ = 0.88 $\pm$ 0.01), whereas HX has higher 6-fold ($\Psi_6$ = 0.90, $\text{Z}_6$ = 0.89) but lower 4-fold ($\Psi_4$ = 0.54 $\pm$ 0.01, $\text{Z}_4$ = 0.77 $\pm$ 0.01) global order parameter values. In both RB and HX, the NC domains do not show any structural ordering and are apparently distinguishable from the C regions in the Voronoi construction (Fig. S2) and Delaunay triangulation (Fig. \ref{fig:System} (C) \& (D), Fig. S3). However, the C regions, \textit{i.e.} the crystallites, are quantitatively identified by distinctly higher values of the local $m$-fold bond-orientational  ($\psi_m$) and positional ($\zeta_m$) order parameters of the constituent particles than those in the NC domains (SM). All particles with $\left| \psi_4\right| \geq$ 0.61 and $\operatorname{Re}(\zeta_4) \geq$ 0.82 are recognized as having RB crystalline ordering, and the local order parameter criteria for identifying particles with HX symmetry are defined as $\left| \psi_6\right|  \geq$ 0.90 and $\operatorname{Re}(\zeta_6) \geq$ 0.90.

In dense systems of anisotropic particles, such as squares, the translational and rotational dynamics of a particle are significantly affected by the configuration of its neighbors owing to interdigitation. Therefore, we considered the order parameter values of the neighbors too in identifying the particles constituting either the C or NC regions. According to our definition, all crystalline particles with more than three crystalline neighbors constitute C domains, and non-crystalline particles with at least three similar neighbors form the NC regions in both RB and HX. Fig. \ref{fig:System} (C) and (D) show the C and NC regions that are identified as per these criteria in RB and HX, respectively. Particles that do not belong to either the C or NC regions were not considered for further analyses. In both the RB and HX, the average area-fractions of the crystallites ($\Phi_{\text{RB-C}}$ = 0.786 $\pm$ 0.001, \& $\Phi_{\text{HX-C}}$ = 0.771 $\pm$ 0.001) are not meaningfully higher than those of the NC domains ($\Phi_{\text{RB-NC}}$ = 0.775 $\pm$ 0.002, \& $\Phi_{\text{HX-NC}}$ = 0.763 $\pm$ 0.002).

\begin{figure*}[t!] 
	\centering
	\includegraphics[width=\linewidth]{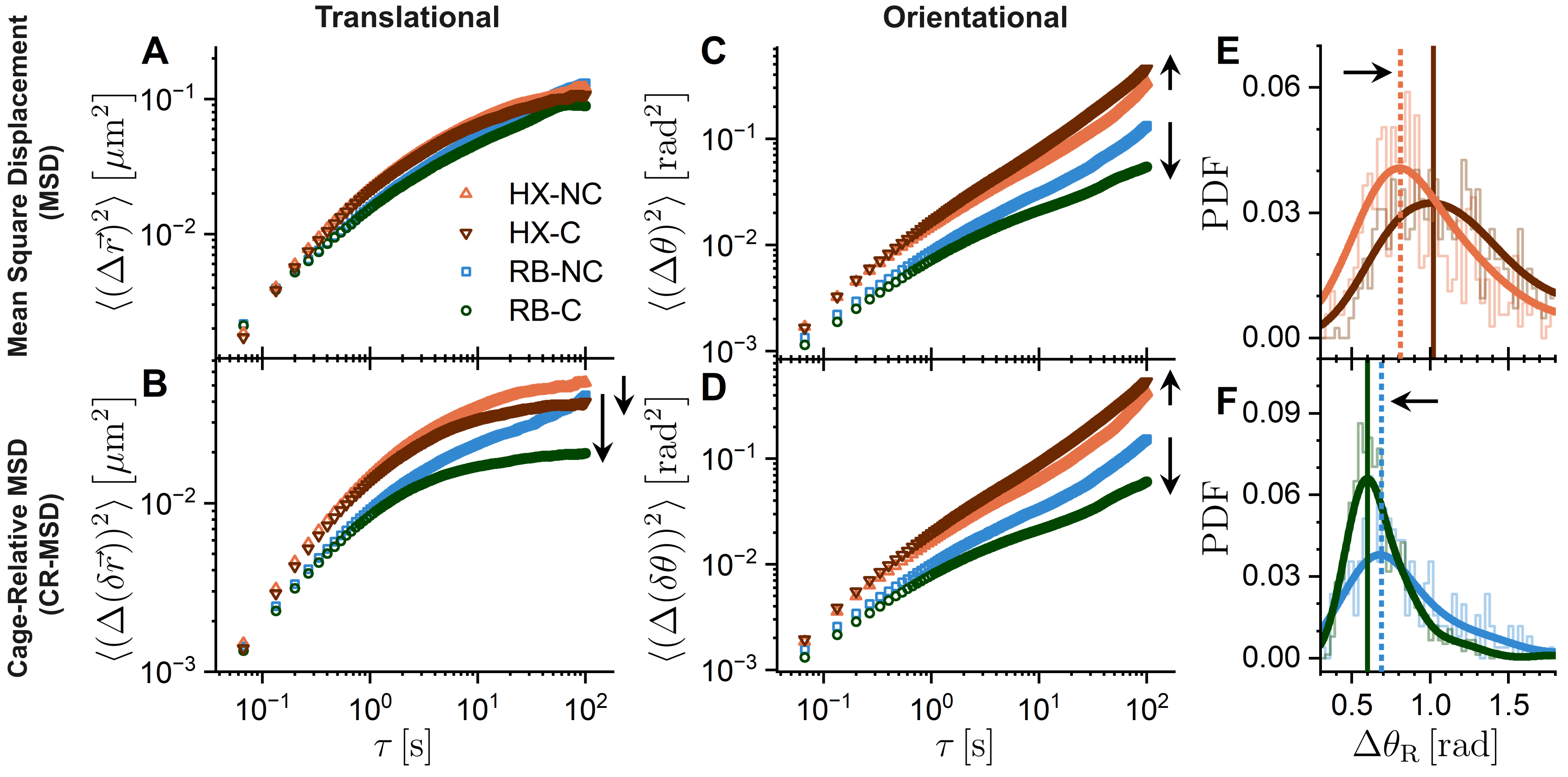}
	\caption{Translational and orientational MSDs, CR-MSDs (see text for definition), and probability distributions of the range of accessible orientational states of the particles over the observation time (\SI{160}{s}) with different structural properties. The ensemble- and time-averaged translational MSDs (A) and CR-MSDs (B) of the particles in HX-NC, HX-C, RB-NC, and RB-C are shown with open triangles, inverted triangles, squares, and circles, respectively. The colors are in accordance with those of the square markers in Fig. \ref{fig:System} (C) \& (D). The corresponding orientational MSDs (C) and CR-MSDs (D) are shown in the second column with the same symbols and colors. Standard errors for the MSDs, which are negligibly small, are not shown in the plots (SM). The normalized probability distributions of the range of accessible orientational states of the particles ($\Delta\theta_{\text{R}}$) belonging to the C and NC domains in the HX (E) and RB (F) are plotted with the same colors. Thick curves are smoothed out of the block-bar plots, and the vertical lines (solid for C domains and dashed for NC regions) mark the positions of the peaks, indicating the most probable values. The arrows indicate the changes in the corresponding quantities upon crystallization.}
	\label{fig:Dynamics}
\end{figure*}

It is important to note that the NC domains are not grain boundaries, which are narrow lines of particles separating crystallites of different orientations. Moreover, the dynamic characteristics of the particles in the NC regions compared with those in C (Fig. \ref{fig:Dynamics}), ensure that they are neither trapped in any local minima of the free energy landscape nor are they arrested dynamically, differentiating the NC regions from grain boundaries. The NC domains are instead approaching crystallization quasi-statically as the particles reorganize to achieve configurations of progressively lower free energy and eventually join neighboring C domains. We also compared $\Psi_m$ and $\text{Z}_m$ values in the NC regions to examine the existence of a hexatic phase \cite{Nelson1979}. However, considerably lower values of both order parameters in RB-NC ($\Psi_4$ = 0.41 $\pm$ 0.01, $\text{Z}_4$ = 0.67 $\pm$ 0.01, and $\Psi_6$ = 0.78 $\pm$ 0.02, $\text{Z}_6$ = 0.80 $\pm$ 0.01) and HX-NC ($\Psi_6$ = 0.80 $\pm$ 0.01, $\text{Z}_6$ = 0.80 $\pm$ 0.01), compared to those in RB-C ($\Psi_4$ = 0.73, $\text{Z}_4$ = 0.97) and HX-C ($\Psi_6$ = 0.95, $\text{Z}_6$ = 0.95), respectively, ensure the absence of any ordering in the NC regions.

\subsection*{Configurational Dynamics}

To identify the crystallization pathways, \textit{i.e.}, the configurational dynamics that drive the structural evolution of the dense assemblies of squares to eventually equilibrate into crystallites of different symmetries at varied osmotic pressures yet very close area fractions, we analyzed both the translational ($\vec{r} (t)$) and rotational ($\theta (t)$) motions of the particles constituting the C and NC regions in the RB and HX (Fig. S4). The translational and rotational mean square displacements (MSD) of the particles, $\left\langle (\Delta \vec{r})^2\right\rangle$ and $\left\langle (\Delta \theta)^2\right\rangle$, are shown in Fig. \ref{fig:Dynamics} (A) \& (C), respectively. Fig. \ref{fig:Dynamics} (B) \& (D) exhibits the corresponding cage-relative MSDs (CR-MSD), which are calculated from the relative displacements of a particle with respect to a cage formed by its neighbors (SM) \cite{Mazoyer2009,Illing2017} and are denoted as $\left\langle (\Delta(\delta\vec{r}))^2\right\rangle$ and $\left\langle (\Delta(\delta\theta))^2\right\rangle$, respectively. These MSDs and CR-MSDs were averaged over all particles constituting each of the four structural phases,\textit{ i.e.}, HX-NC, HX-C, RB-NC, and RB-C. The trends of these MSDs reveal the characteristic reorganization dynamics of the particles, leading to either RB or HX ordering.

All translational MSDs largely overlap, the ones for HX-NC and HX-C being slightly higher at intermediate time-lags (Fig. \ref{fig:Dynamics} (A)). However, CR-MSDs, which are consistently smaller than the corresponding MSDs, deviate from each other with increasing time-lag, uncovering single-particle dynamics with respect to its neighbors. Translational CR-MSDs for RB are lower than those for HX, while CR-MSDs diminish upon crystallization for both RB and HX (Fig. \ref{fig:Dynamics} (B)). Thus, the positional displacements of the particles relative to their neighbors decrease consistently with an infinitesimal increase in concentration. Moreover, a considerable difference between the MSDs and their cage-relative counterparts reveals the presence of long-range Mermin-Wagner translational fluctuations \cite{Illing2017}, which is dominant in RB-C, where the particles are mostly side-aligned. The long-range collective positional displacements progressively diminish with the orientational disorder of the squares, closing the gap between the corresponding translational CR-MSD and MSD. Hence, the CR-MSDs are larger,\textit{ i.e.}, closer to the corresponding MSDs, for HX-C and HX-NC at all time-lags, whereas for RB-NC, the CR-MSD increases sharply at longer times, where it becomes comparable to those for HX-C and HX-NC.

In the case of rotational dynamics, CR-MSDs are similar to MSDs for all configurations over the entire time-lag range, confirming the absence of any collective rotational motion of the particles, even at short times (Fig. \ref{fig:Dynamics} (C) \& (D)). The MSDs for RB are lower than those for HX, and they decrease upon crystallization in RB, following a similar trend to translational CR-MSDs. Intriguingly, the rotational dynamics of the particles in the HX exhibit anomalous variation. The MSD for HX-C is larger than that for HX-NC at longer time-lag values, implying wider orientational fluctuations of the particles as they crystallize into HX.

\subsection*{Range of Accessible Orientational States}

The angular spread of the orientational fluctuations of a particle within the observation time is given by the kinematic parameter $\Delta\theta_{\text{R}} = (\theta(t)_{\text{max}} - \theta(t)_{\text{min}})$, where $\theta(t)_{\text{max}}$ and $\theta(t)_{\text{min}}$ are the maximum and minimum values in the orientation time series $\theta(t)$, respectively. Therefore, the value of $\Delta\theta_{\text{R}}$ increases monotonically with the observation duration as the rotational fluctuations become wider and it eventually saturates when the particle passes through the entire orientational space available within the dynamic confinement imposed by its neighbors. For a slowly evolving system, such as ours, the dynamic confinement changes gradually, allowing most of the particles to explore all accessible orientational states. Consequently, $\Delta\theta_{\text{R}}$ represents the range of accessible orientational microstates to a particle and thus provides a direct measure of orientational entropy. Furthermore, the value of $\Delta\theta_{\text{R}}$ for one particle may vary from that of another in the same structural domain depending on the configuration of their neighbors, resulting in a finite distribution of $\Delta\theta_{\text{R}}$. We plotted the normalized probability distributions of $\Delta\theta_{\text{R}}$ for the particles constituting HX-NC \& HX-C (Fig. \ref{fig:Dynamics} (E)), and RB-NC \& RB-C (Fig. \ref{fig:Dynamics} (F)). The peak of the smoothed probability distribution indicates the most probable value of $\Delta\theta_{\text{R}}$ in the corresponding structural phase. Therefore, it is evident that $\Delta\theta_{\text{R}}$ for the majority of the particles increases considerably, from 0.81 rad to 1.02 rad, \textit{i.e.}, $\approx 25\%$, as the system crystallizes to HX, even though $\Phi_{\text{HX-C}}$ is slightly higher than $\Phi_{\text{HX-NC}}$. In RB, the most probable value of $\Delta\theta_{\text{R}}$ decreases, from 0.69 rad to 0.60 rad, upon crystallization.

It is important to note that as the most-probable value of $\Delta\theta_{\text{R}}$ attains saturation, signifying that most of the particles have explored their accessible orientational microstates, the orientational MSD may still increase because of the contributions from a few particles that perform progressively wider orientational motion allowed by the configuration of their neighbors. This is also evident from the extended tails of the probability distribution of $\Delta\theta_{\text{R}}$. Thus, compared with the orientational MSD, the most-probable value of $\Delta\theta_{\text{R}}$ provides a more reliable measure of the accessible orientational states and, hence, the orientational entropy of the particles in a structural domain.

\begin{figure*}[!h]
	\centering
	\includegraphics[width=1\linewidth]{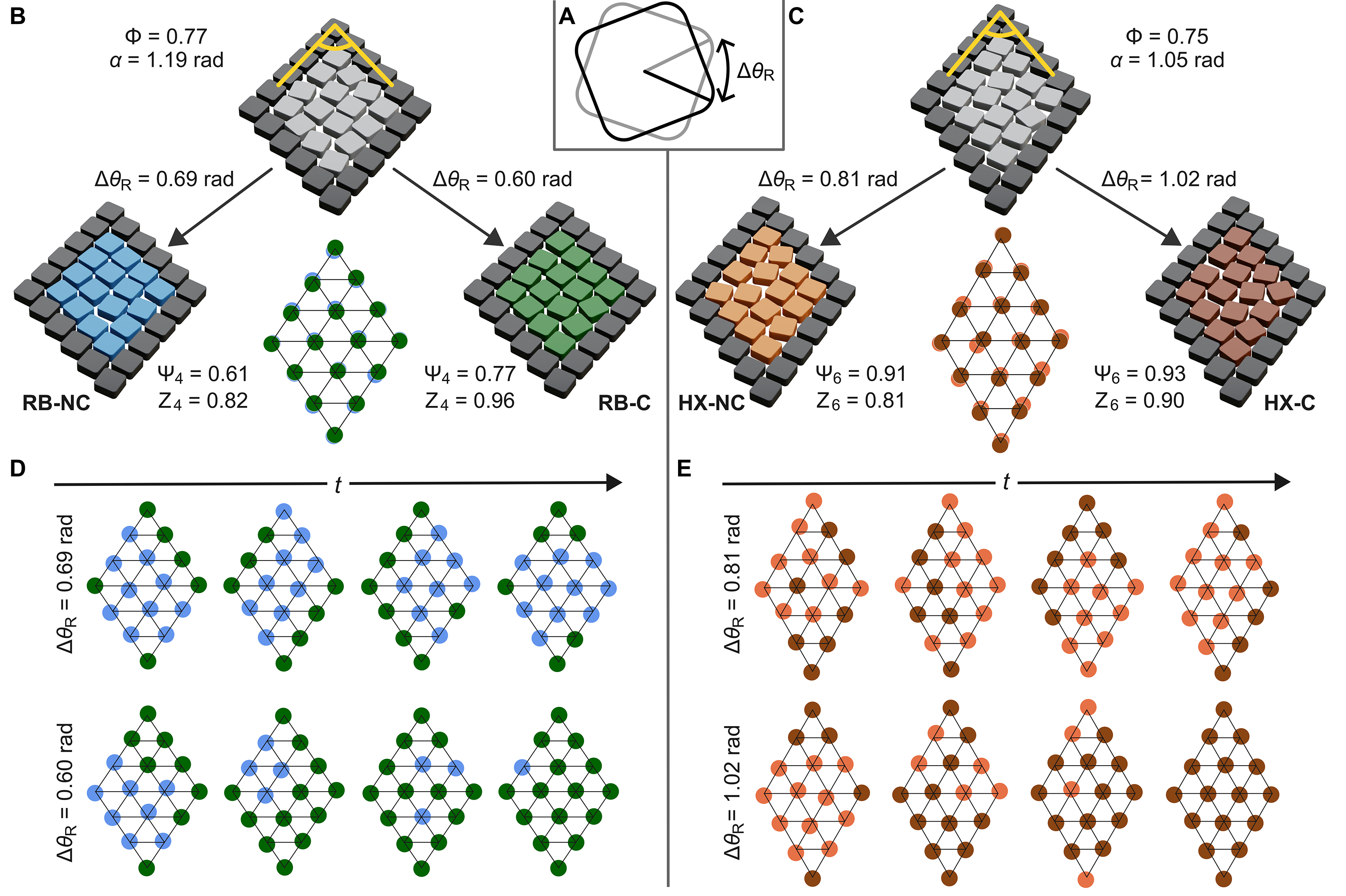}
	\caption{Simulated structural evolution and equilibrium phases of hard square self-assemblies with various ranges of accessible rotational states, $\Delta\theta_{\text{R}}$. (A) The rotational motion of each particle is confined within a range $\Delta\theta_{\text{R}}$, which is the angle between two extreme orientations (black and grey squares) that can be attained as the system equilibrates. A set of sixteen corner-rounded square platelets (light grey), confined by twenty static boundary particles (dark grey) at a specific area fraction ($\Phi$) and lattice angle ($\alpha$) corresponding to their observed values in RB (B) and HX (C), are allowed to reorganize through Brownian diffusion without overlapping. Each initial configuration is equilibrated with two different values of  $\Delta\theta_{\text{R}}$ as observed in the experiments (Fig. \ref{fig:Dynamics} (E) \& (F)). Global 4-fold bond-orientational ($\Psi_4$) and positional ($\text{Z}_4$) order parameter values, considering four central particles, identify the equilibrated structures as RB-NC ($\Psi_{4}$ = 0.61 $\pm$ 0.12, $Z_{4}$ = 0.82 $\pm$ 0.06) and RB-C ($\Psi_{4}$ = 0.77 $\pm$ 0.02, $Z_{4}$ = 0.96 $\pm$ 0.01). Similarly, the simulated configurations in the HX are recognized as HX-NC ($\Psi_{6}$ = 0.91 $\pm$ 0.04, $Z_{6}$ = 0.81 $\pm$ 0.05) and HX-C ($\Psi_{6}$ = 0.93 $\pm$ 0.02, $Z_{6}$ = 0.90 $\pm$ 0.02) by their $\Psi_6$ and $\text{Z}_6$ values. All equilibrium configurations are shown using colors that are consistent with previous figures. In the middle of each side (B \& C), the corresponding perfect lattice triangulations (black lines) are overlaid on equilibrated particle centers, which are shown as filled circles with the same colors, to demonstrate emergent structural ordering pictorially. Typical equilibration pathways are depicted by four intermediate structural configurations with progressing times for each case in the middle rows (D \& E). Color-coded filled circles mark the centers of the squares, and the colors denote the local structural ordering defined by the values of the local order parameters $\psi_m$ and $\zeta_m$.}
	\label{fig:Simulation}
\end{figure*}

To investigate this intriguing trend of $\Delta\theta_{\text{R}}$ and its influence on the reorganization dynamics of the particles and the subsequent structural evolution leading to crystallization, we performed Brownian dynamics (BD) simulations on smaller sizes of the same systems. A small system size was chosen to examine the equilibration of the local confined structural domains. We considered 36 hard-interacting squares of the same size and corner rounding as in the real system. The static boundary particles set the area fraction ($\Phi$) and lattice angle ($\alpha$) to resemble the RB and HX neighborhoods. The inner 16 particles were allowed to attain accessible configurations through translational and rotational diffusion, avoiding overlapping and complying with a constrained range of their orientational fluctuations, as per the corresponding observed values of $\Delta\theta_{\text{R}}$. Thus, the maximum rotational states accessible to each mobile particle were restricted to a predefined range given by $\Delta\theta_{\text{R}}$ (Fig. \ref{fig:Simulation} (A)). For each RB and HX, a randomly chosen disordered initial configuration was allowed to equilibrate with two different values of $\Delta\theta_{\text{R}}$ corresponding to its most probable values in C and NC (Fig. \ref{fig:Dynamics} (E) \& (F)), in two separate simulation runs. The symmetries of the four final equilibrated configurations are defined by their global order parameters, $\Psi_m$ and $\text{Z}_m$. The simulation results are presented in Fig. \ref{fig:Simulation} (B) \& (C); they completely agree with the experimental observations and corroborate our inferences. In the case of RB, the system with a smaller $\Delta\theta_{\text{R}}$ (0.60 rad) crystallizes, whereas the system with a larger $\Delta\theta_{\text{R}}$ (0.69 rad) but the same area fraction does not equilibrate into an ordered structure. As observed in the experiments, the trend is opposite in the case of HX. The system with $\Delta\theta_{\text{R}} = $1.02 rad shows hexagonal ordering but does not crystallize at the same area fraction when $\Delta\theta_{\text{R}}$ is narrower (0.81 rad). Moreover, the trends of the translational and rotational CR-MSDs (calculated from the simulated dynamics of the four inner particles) too are in good agreement with those from the experiments (Fig. S6). The intermediate configurations at progressing times depict the structural evolution, illustrating the pathways of crystallization as the systems approach equilibrium with different permissible values of $\Delta\theta_{\text{R}}$ (Fig. \ref{fig:Simulation} (D) \& (E)). This pictorial illustration decisively demonstrates how a decrease or increase in the range of accessible orientational states of the squares dictates the route to crystallization, eventually leading to two distinctly different symmetry structures.

\begin{figure*}[t!]
	\centering
	\includegraphics[width=0.60\linewidth]{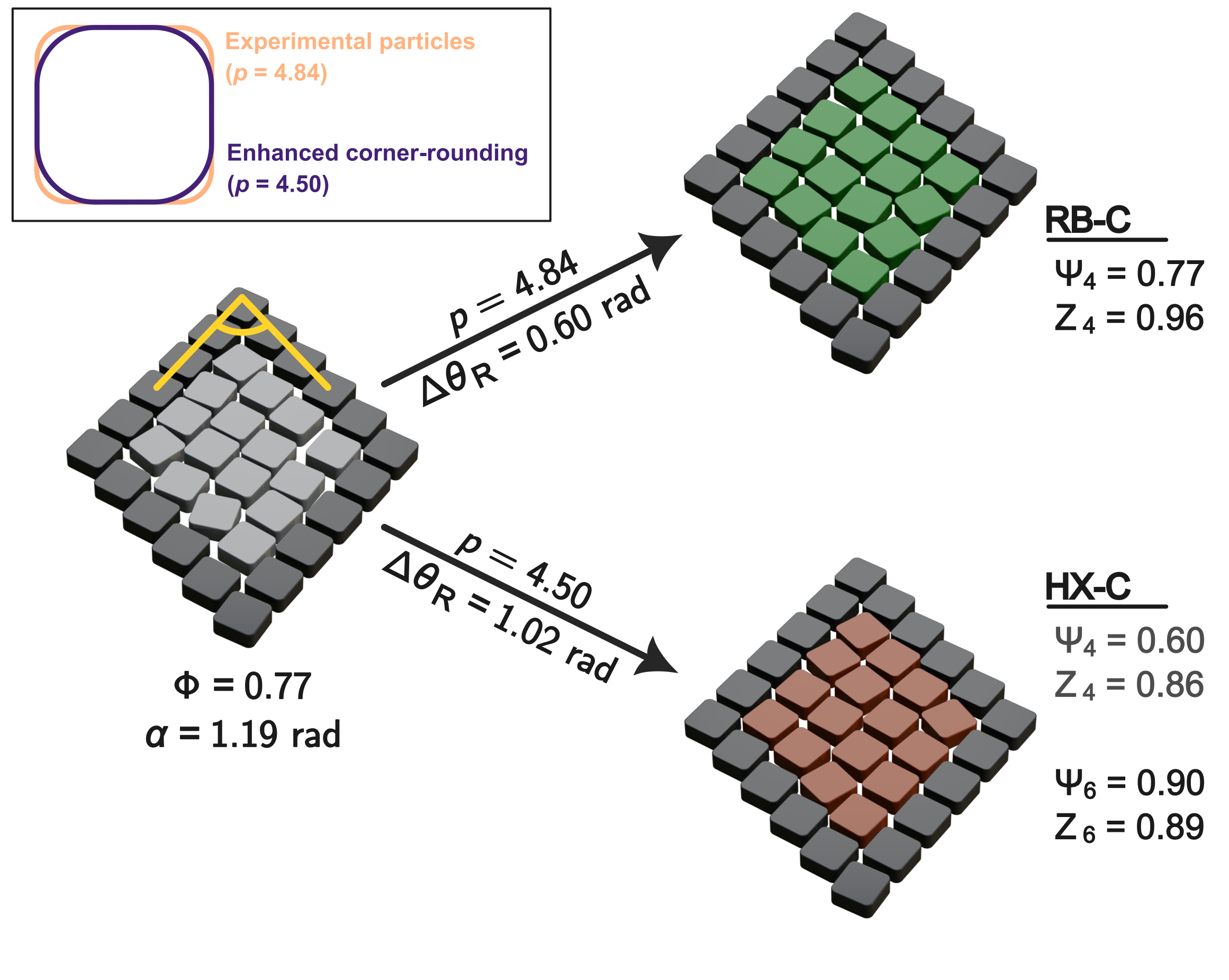}
	\caption{BD simulation of the crystallization of the same initial configuration of square particles with two different values of $\Delta \theta_{\text{R}}$. An initial random RB configuration with the same $\Phi$ and $\alpha$ values as in Fig. \ref{fig:Simulation}(B), is equilibrated with different values of $\Delta \theta_{\text{R}}$, 0.60 rad and 1.02 rad, corresponding to RB-C and HX-C, as shown in Fig. \ref{fig:Dynamics}(E) and (F), respectively. Global 4-fold and 6-fold positional and bond-orientational order parameters calculated over four central particles recognize the symmetry of equilibrated structural ordering as RB-C ($\Psi_{4}$ = 0.77 $\pm$ 0.02, $Z_{4}$ = 0.96 $\pm$ 0.01) and HX-C ($\Psi_{4}$ = 0.60 $\pm$ 0.07, $Z_{4}$ = 0.86 $\pm$ 0.02, $\Psi_{6}$ = 0.90 $\pm$ 0.02, $Z_{6}$ = 0.89 $\pm$ 0.03). The inset shows two squares with different degrees of corner rounding defined by the value of $p$, which is decreased from 4.84, corresponding to the experimental particles, to 4.50, allowing a wider $\Delta \theta_{\text{R}}$ (1.02 rad) at the higher density ($\Phi$ = 0.77).}
	\label{fig:WiderRange}
\end{figure*}

\subsection*{Indirect Effect of Density}

To further explore the explicit effects of $\Delta\theta_{\text{R}}$ in comparison with that of the area fraction ($\Phi$) and neighboring lattice angle ($\alpha$) on the configurational dynamics of the particles and their subsequent crystallization into a preferred symmetry structure, we varied $\Delta\theta_{\text{R}}$ while keeping $\Phi$ and $\alpha$ unchanged. This was achieved by enhancing the corner rounding of the square particles (Fig. \ref{fig:WiderRange} inset), thereby circumventing the restriction imposed by the area fraction on the upper limit of $\Delta\theta_{\text{R}}$. Implementing this in a similar simulation, the same initial configuration of RB, \textit{i.e.}, defined by its $\Phi$ (= 0.77) and $\alpha$ (= 1.19 rad) (as in Fig. \ref{fig:Simulation} (B)) was allowed to equilibrate with two different values of $\Delta\theta_{\text{R}}$, 0.60 rad and 1.02 rad corresponding to that for RB-C and HX-C, respectively. Intriguingly, the same system that crystallizes to RB-C with $\Delta\theta_{\text{R}}$ = 0.60 rad, equilibrates to HX-C when the particles can access a wider range of orientational states, given by $\Delta\theta_{\text{R}}$ = 1.02 rad, as shown in Fig. \ref{fig:WiderRange}. The equilibrated hexagonal ordering is verified by lower values of 4-fold global order parameters ($\Psi_4$ = 0.60, \& $\text{Z}_4$ = 0.86) and higher values of their 6-fold counterparts ($\Psi_6$ = 0.90, \& $\text{Z}_6$ = 0.89), similar to those for HX-C. The enhancement in corner rounding reduced the value of $\Phi$ marginally ($\approx$ 0.7\%), which can be neglected for all practical purposes.

In a real dense system of squares, variations in the 2D osmotic pressure ($\Pi$), area fraction ($\Phi$), and $\Delta\theta_{\text{R}}$ are coupled in an intricate manner. However, in our simulations, we decoupled these dependencies and varied $\Delta\theta_{\text{R}}$ independently while keeping $\Phi$, $\alpha$, and $\Pi$ unchanged in different simulation runs. Therefore, our results unequivocally conclude that $\Delta\theta_{\text{R}}$ solely governs the configurational dynamics of the squares, and thus, the pathways of crystallization into the optimal equilibrium ordering, where $\Phi$ and $\alpha$ indirectly influence the process by setting an upper bound of $\Delta\theta_{\text{R}}$.

\subsection*{Free Energy Landscape}

We derived the variation in free energy as a function of $\Delta\theta_{\text{R}}$ to discern the effect of $\Delta\theta_{\text{R}}$ on the crystallization pathways leading to the preferred equilibrium ordering and the cause of an apparent anomalous increase in the accessible orientational states of the squares when the system crystallized into HX-C. The change in the Gibbs free energy ($\Delta G = \Delta F + \Delta(\Pi A)$) is practically the same as that in the Helmholtz free energy ($\Delta F$) for crystallization of the RB-NC and HX-NC regions into RB-C and HX-C, respectively, because the ordering is achieved through structural reconfigurations at a constant 2D osmotic pressure ($\Pi$) and with an insignificant change in $\Phi$, and consequently, in area ($A$)  (SM). For this system of hard-interacting particles, $\Delta F$ varies solely because of the change in the configuration entropy ($S$) as $\Delta F = - T \Delta S$. Therefore, for anisotropic particles, $\Delta F$ consists of both translational ($\Delta F_{\text{t}}$) and rotational ($\Delta F_{\text{r}}$) contributions, \textit{i.e.}, $\Delta F = \Delta F_{\text{t}} + \Delta F_{\text{r}} $, resulting from corresponding configurational variations.

\begin{figure}[!h]
	\centering
	\includegraphics[width=0.5\linewidth]{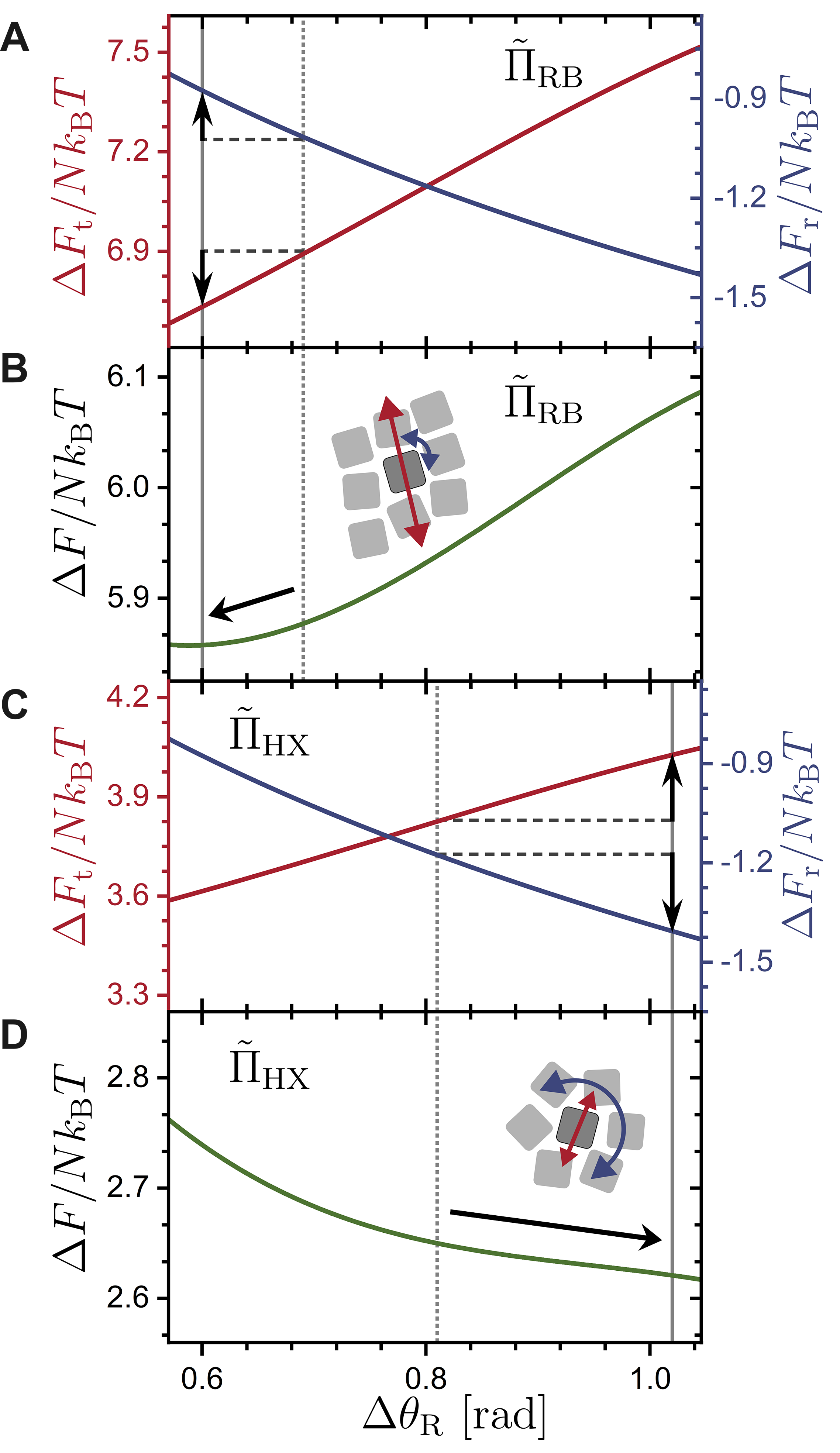}
	\caption{The variation in the free energy ($\Delta F/Nk_{\text{B}}T$), translational ($\Delta F_{\text{t}}/Nk_{\text{B}}T$), and rotational ($\Delta F_{\text{r}}/Nk_{\text{B}}T$) contributions therein, with $\Delta\theta_{\text{R}}$. $\Delta F_{\text{t}}/Nk_{\text{B}}T$ (dark red) and $\Delta F_{\text{r}}/Nk_{\text{B}}T$ (blue) are plotted against $\Delta\theta_{\text{R}}$ using the left and right axes (same scale), respectively, for $\tilde{\Pi}_{\text{RB}} = 5.77$ (A) and $\tilde{\Pi}_{\text{HX}} = 3.12$ (C). The corresponding resultant free-energy variations $\Delta F/Nk_{\text{B}}T$ (green) are shown in (B) and (D), respectively. The vertical lines mark the $\Delta\theta_{\text{R}}$ values for the NC regions (dotted lines) and after crystallization (solid lines) for both RB and HX. The black arrows indicate the change in the free energy values upon crystallization. The insets in (B) and (D) are cartoon representations of RB and HX structures, where overlaid color-coded arrows exhibit the relative contributions of accessible translational and orientational states of a particle (dark grey) in minimizing the respective configurational free energies.}
	\label{fig:FreeEnergy}
\end{figure}

The translational component $F_{\text{t}}$ is related to $\Pi$, and $\Phi$ ($= A_{\text{p}}N/A$, where $N$ particles, each with area $A_{\text{p}}$, occupy the total area $A$) by $\Pi = - \left.\frac{\partial F_{\text{t}}}{\partial A} \right|_{N}$ \cite{Smallenburg2012,Thorneywork2017}.  Thus, $ F_{\text{t}}$ can be derived from $\Pi(\Phi)$ as $F_{\text{t}} = NA_{\text{p}}\int (\Pi(\Phi)/\Phi^{2})\,d\Phi$. For this osmotically compressed 2D assembly of squares on an inclined plane, both $\Pi$ and $\Phi$ increase monotonically towards the lower end. Their implicit relationship is given by $\tilde{\Pi}(\Phi) = \Pi (\Phi)/(k_{\text{B}}T/A_{\text{p}}) = \Phi^* \ln (\Phi^*/(\Phi^* - \Phi))$, where $\Phi^*$ corresponds to the densest packing at the bottom end \cite{Wang2015,Hodson2016}. We replaced $\Phi^*$ with $\Phi_{\text{max}} (\Delta\theta_{\text{R}})$, which is the maximum possible value of $\Phi$ for the squares at a given $\Delta\theta_{\text{R}}$, and obtained the variation in translational free energy as $\Delta F_{\text{t}}/Nk_{\text{B}}T = \ln\left[ \exp\left( \tilde{\Pi}/\Phi_{\text{max}}(\Delta\theta_{\text{R}})\right ) -1\right]$ (SM). The rotational contribution, $F_{\text{r}}$, depends logarithmically on the available orientational states, \textit{i.e.}, $\Delta\theta_{\text{R}}$, and is expressed as, $\Delta F_{\text{r}}/Nk_{\text{B}}T = - \ln\left[4 \Delta\theta_{\text{R}}\right]$ \cite{Zhao2011}. Combining these two contributions, we finally had the variation of the resultant dimensionless free energy per particle, $\Delta F/Nk_{\text{B}}T$, with $\Delta\theta_{\text{R}}$ as
\begin{equation}
	\label{Ftot}
	\frac{\Delta F}{Nk_{\text{B}}T} = \ln\left[ \exp\left( \tilde{\Pi}/\Phi_{\text{max}}(\Delta\theta_{\text{\text{R}}})\right) -1\right] - \ln\left[4 \Delta\theta_{\text{R}}\right].
\end{equation}

The value of $\Phi^*$ was obtained empirically by fitting the observed variation in $\Phi$ along the inclined plane (Fig. S7) \cite{Wang2015}. This enabled us to use the expression for  $\tilde{\Pi}(\Phi)$ to calculate the values of $\tilde{\Pi}$ corresponding to RB ($\tilde{\Pi}_{\text{RB}}$ = 5.77) and HX ($\tilde{\Pi}_{\text{HX}}$ = 3.12) from their respective area fractions, $\Phi_{\text{RB}}$ and $\Phi_{\text{HX}}$ (SM). We derived the dependence of $\Phi_{\text{max}}$ on $\Delta\theta_{\text{\text{R}}}$ numerically (SM, Fig. S8) to compute the variation in the dimensionless free energy per particle $\Delta F / Nk_{\text{B}}T$ and its components, $\Delta F_{\text{t}} / Nk_{\text{B}}T$ and $\Delta F_{\text{r}} / Nk_{\text{B}}T$, with $\Delta\theta_{\text{\text{R}}}$ following Eq. \ref{Ftot}, at $\tilde{\Pi}_{\text{RB}}$ and $\tilde{\Pi}_{\text{HX}}$ (Fig. \ref{fig:FreeEnergy}).

At $\tilde{\Pi}_{\text{RB}}$, a decrease in the translational contribution, $\Delta F_{\text{t}} /Nk_{\text{B}}T$, minimizes the resultant configurational free energy, $\Delta F/Nk_{\text{B}}T$, overcompensating for an increase in $\Delta F_{\text{r}} /Nk_{\text{B}}T$ as $\Delta\theta_{\text{R}}$ becomes narrower from its value in RB-NC to that in RB-C. The squares in the RB are mostly side-aligned and can access more translational states, thus maximizing the configuration entropy, even with reduced orientational fluctuations upon crystallization (Fig. \ref{fig:FreeEnergy} (A) \& (B)). By contrast, at a lower osmotic pressure, $\tilde{\Pi}_{\text{HX}}$, a decrease in the rotational free energy, $\Delta F_{\text{r}} /Nk_{\text{B}}T$, overcomes the increase in $\Delta F_{\text{t}} /Nk_{\text{B}}T$ to minimize the total free energy, $\Delta F/Nk_{\text{B}}T$. An increase in $\Delta\theta_{\text{R}}$ allows the particles to access more orientational states, thus minimizing the configurational free energy as the system crystallizes into HX-C, where the translational dynamics of the densely packed misaligned squares are considerably hindered (Fig. \ref{fig:FreeEnergy} (C) \& (D)).

Therefore, it is evident that the structural evolution of the squares towards the minimization of configurational free energy and eventual crystallization into RB or HX is driven by a decrease or increase in $\Delta\theta_{\text{R}}$ at a higher ($\tilde{\Pi}_{\text{RB}}$) or lower ($\tilde{\Pi}_{\text{HX}}$) osmotic pressure, respectively, thus conclusively validating our inference. Moreover, the emergence of both free energy minima at the experimentally observed values of $\Delta\theta_{\text{R}}$ corresponding to RB-C and HX-C quantitatively verifies that the range of accessible orientational states governs the pathways of crystallization into the optimal symmetry structure at a given osmotic pressure.

The free energy variation in the CE phase, which is observed at an intermediate osmotic pressure ($\tilde{\Pi}_{\text{CE}}$) between $\tilde{\Pi}_{\text{RB}}$ and $\tilde{\Pi}_{\text{HX}}$, exhibits two minima around a broad local peak (Fig. S9). Thus, at $\tilde{\Pi}_{\text{CE}}$, even a weak local perturbation in the configurational dynamics of the squares pushes the neighborhood to either the left (smaller $\Delta\theta_{\text{R}}$) or the right (larger $\Delta\theta_{\text{R}}$) minimum to form RB or HX crystallites, respectively. This explains the emergence of a coexistence phase \cite{Zhao2011}, where different $\Delta\theta_{\text{R}}$ values facilitate the formation and stabilization of local RB and HX crystallites at the same area fraction and osmotic pressure, further corroborating our conclusion.

\begin{figure}[!htb]
	\centering
	\includegraphics[width=0.65\linewidth]{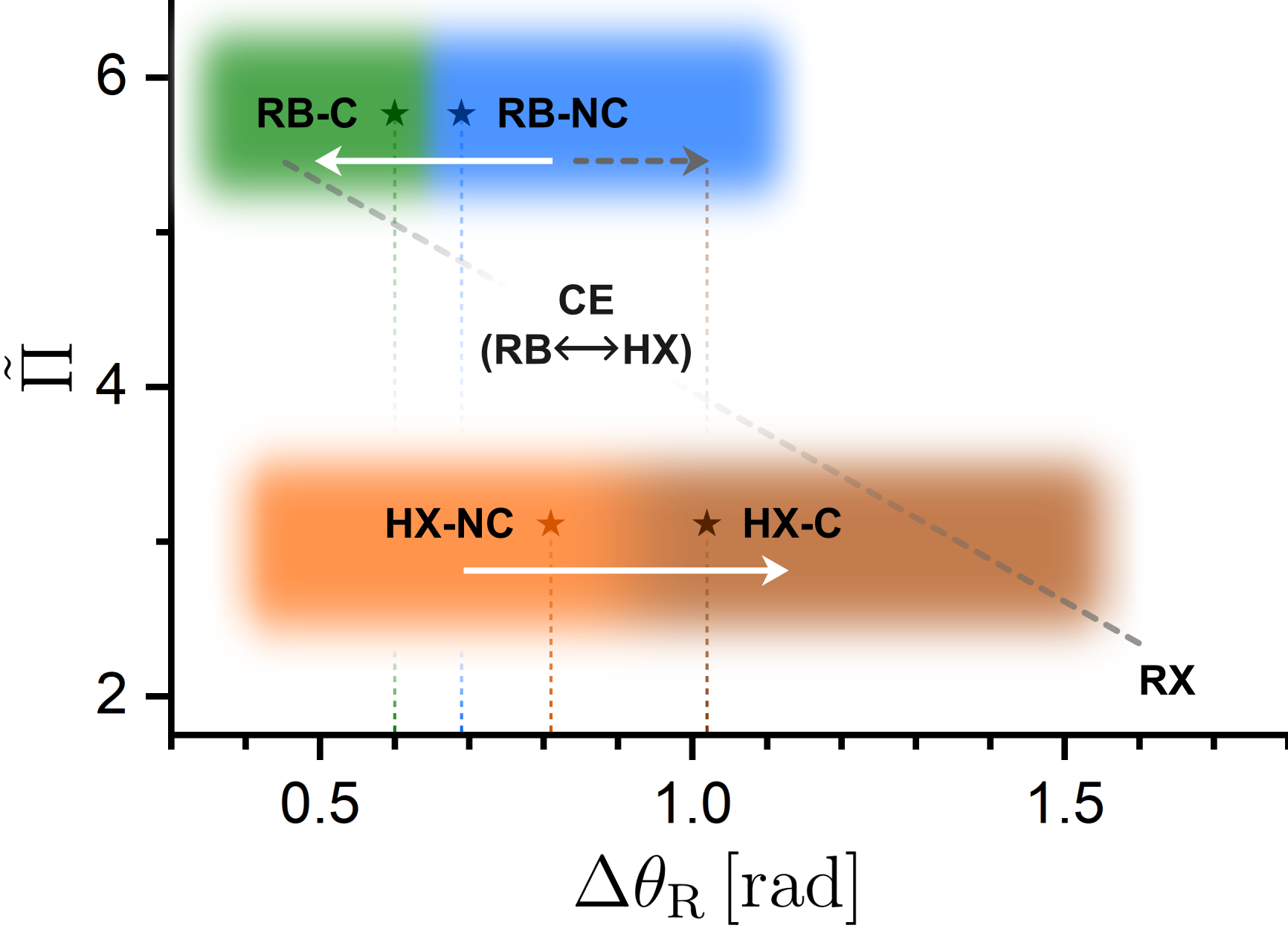}
	\caption{Crystallization pathways in the $\tilde{\Pi}$ - $\Delta\theta_{\text{R}}$ space. Different structural phases and metastable domains formed by a dense system of rounded squares, \textit{i.e.}, RB-NC, RB-C, HX-NC, HX-C, RX, and the coexistence (CE) of RB and HX or RX, are shown in a $\tilde{\Pi}$ - $\Delta\theta_{\text{R}}$ plot. The colored rectangles with smooth boundaries corresponding to RB-C, RB-NC, HX-C, and HX-NC represent the areas over which they are likely to form, whereas filled color-coded stars denote their exact locations, as observed in the experiments. A crystal-crystal transition under varied osmotic pressure \cite{Zhao2011} is shown by the dashed diagonal line. While the white horizontal arrows denote the routes of crystallization from unstable points (RB-NC and HX-NC) on the free energy landscape (Fig. \ref{fig:FreeEnergy}) to the corresponding minima (RB-C and HX-C), the grey horizontal arrow indicates the transition from RB-NC to HX-C, where greater corner rounding of the squares allows a wider $\Delta\theta_{\text{R}}$ at $\tilde{\Pi}_{\text{RB}}$ (Fig. \ref{fig:WiderRange}).}
	\label{fig:PhaseDiagram}
\end{figure}

\subsection*{Crystallization Pathways}

The routes of crystallization in a dense system of Brownian squares, \textit{i.e.}, the pathways of equilibration of the NC structural domains into the respective C phases, are pictorially shown in the $\tilde{\Pi}$ - $\Delta\theta_{\text{R}}$ space in Fig. \ref{fig:PhaseDiagram}. $\Delta\theta_{\text{R}}$, which provides a direct measure of the orientational entropy per particle, solely governs the crystallization pathways and, hence, is used as the independent parameter. The horizontal arrows denote the routes of crystallization with a variation in $\Delta\theta_{\text{R}}$ at a given osmotic pressure, $\tilde{\Pi}$. When a dense system of rounded Brownian squares attains equilibrium under osmotic compression, a continuous crystal-crystal transition is observed at varied $\tilde{\Pi}$ \cite{Zhao2011}, accompanied by a corresponding variation in $\Delta\theta_{\text{R}}$, as indicated by the dashed diagonal line. While the system equilibrates at a higher $\tilde{\Pi}$, around $\tilde{\Pi}_{\text{RB}}$, the range of the accessible orientational states of the squares ($\Delta\theta_{\text{R}}$) reduces to drive the system from RB-NC, which is unstable in the free energy landscape, to the free energy minimum, corresponding to RB-C ordering (Fig. \ref{fig:FreeEnergy}(B)), along the leftward horizontal white arrow. In the region with lower $\tilde{\Pi}$, around $\tilde{\Pi}_{\text{HX}}$, the equilibration of the system is driven by an increase in $\Delta\theta_{\text{R}}$ as HX-C ordering is attained from HX-NC to minimize the free energy (Fig. \ref{fig:FreeEnergy}(D)), as indicated by the rightward horizontal white arrow.

With an appropriately increased corner rounding of the squares, the particles can access a wider range of orientational states at the same osmotic pressure and area fraction as those corresponding to RB. This modifies $\Phi_{\text{max}}(\Delta \theta_{\text{R}})$, and thereby the free energy curve, where the minimum shifts to a larger value of $\Delta\theta_{\text{R}}$, corresponding to its value in HX-C. Consequently, an increase in $\Delta\theta_{\text{R}}$ drives the system to attain HX-C ordering, even at a higher $\tilde{\Pi} \approx \tilde{\Pi}_{\text{RB}}$, which is indicated by the horizontal grey dashed arrow in the phase diagram, and is demonstrated in Fig. \ref{fig:WiderRange}. Therefore, while transitions from one phase to another can be achieved in experiments by controlling $\tilde{\Pi}$ and $\Phi$, thus indirectly regulating $\Delta\theta_{\text{R}}$, the underlying kinetic pathways of the transitions are entirely dictated by $\Delta\theta_{\text{R}}$.

\begin{figure}[!thbp]
	\centering
	\includegraphics[width=1\linewidth]{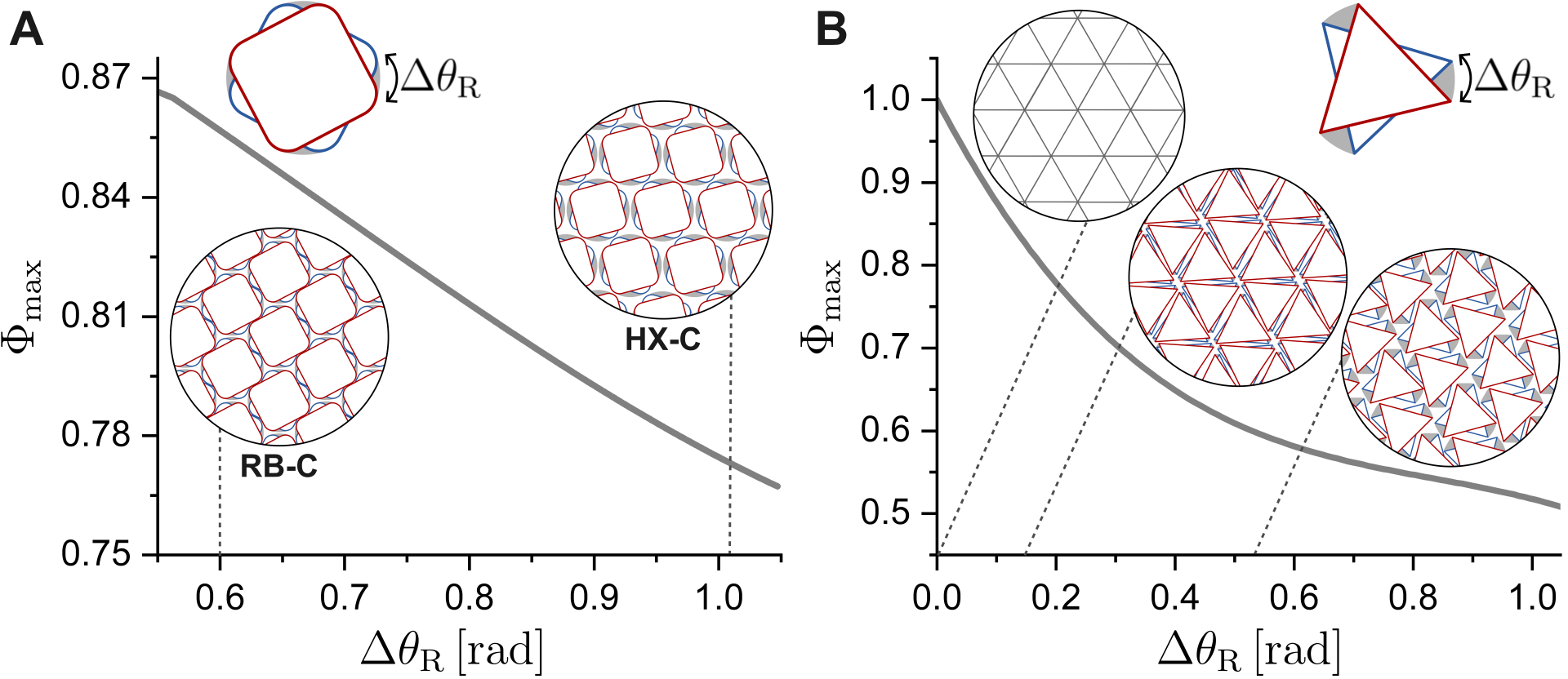}
	\caption{Generality of the monotonic decrease in $\Phi_{\textrm{max}}$ with $\Delta\theta_\textrm{R}$. The numerically simulated values of $\Phi_{\textrm{max}}$ at varying $\Delta\theta_\textrm{R}$ are shown for the dense self-assembly of two convex anisotropically shaped particles, such as squares (A) and triangles (B). The two extreme orientational states accessible by the particles are shown with blue and red boundaries to pictorially define $\Delta\theta_\textrm{R}$, where the rotational excluded areas swept by the corners are marked with grey shading (insets). The densest packing configurations, exhibiting varied structural symmetries, for a few characteristic values of $\Delta\theta_\textrm{R}$ (pointed by grey dashed lines) are shown as circular insets. (A) For the system of squares, RB-C and HX-C configurations are shown at their corresponding $\Delta\theta_\textrm{R}$ values. (B) The simulated configurations of Brownian triangles at the densest packing, which fully agree with experimental observations \cite{Zhao2012a}, are exhibited for three different values of $\Delta\theta_\textrm{R}$.}
	\label{fig:Generality}
\end{figure}

\subsection*{Applicability to Similar Systems}

Minimization of the free energy or maximization of the configurational entropy with $\Delta\theta_{\text{R}}$ to achieve the preferred equilibrium ordering through competition between the translational and orientational contributions, as quantitatively described by our free energy equation (Eq. \ref{Ftot}), is generic in nature. The monotonic decrease in the rotational component of the free energy ($\Delta F_{\text{r}} /Nk_{\text{B}}T$) with $\Delta\theta_{\text{R}}$ is independent of the shape of the anisotropic particle, whereas the exact variation in its translational counterpart ($\Delta F_{\text{t}} /Nk_{\text{B}}T$) depends on the shape of the constituent. However, for any anisotropic convex particle shape, $\Delta F_{\text{t}} /Nk_{\text{B}}T$ increases progressively with $\Delta\theta_{\text{R}}$ because $\Phi_{\text{max}}$ ($\Delta\theta_{\text{R}}$), which represents the theoretical maximum value of $\Phi$ corresponding to a reference state where the particles have only adequate accessible free space to allow orientational fluctuations over the range $\Delta\theta_{\text{R}}$, in general, is a monotonically decreasing function (SM).

To support this generality, we computed the dependence of $\Phi_{\text{max}}$ on $\Delta\theta_{\text{R}}$ for a similar system, \textit{e.g.}, triangular platelet particles \cite{Zhao2012a}, using the same numerical simulation as that used for square particles (Fig. S8). Both systems of square and triangular particles exhibit a similar gradual decrease in $\Phi_{\text{max}} (\Delta\theta_{\text{R}})$ as shown in Fig. \ref{fig:Generality}, where the simulated structural configurations at the densest area fraction for various values of $\Delta\theta_{\text{R}}$ completely agree with the experimental observations in both cases (Fig. \ref{fig:Generality} insets). The monotonically decreasing variation in $\Phi_{\text{max}} (\Delta\theta_{\text{R}})$ explicitly verifies the progressive growth in the translational free energy with $\Delta\theta_{\text{R}}$ (Eq. \ref{Ftot}) for both the shapes, as well as for other anisotropic convex shapes.

The opposing trends of the two free energy components ensure the emergence of at least one boundary or local minimum in the total free energy at a certain value of $\Delta\theta_{\text{R}}$. Based on the specific variations in the translational and rotational contributions, there can be more than one minimum, signifying a coexistence phase, as has been demonstrated for square particles (Fig. S9). While the position of the minimum defines the value of $\Delta\theta_{\text{R}}$ corresponding to the equilibrium ordering, the free-energy variation around the minimum describes the direction of change in $\Delta\theta_{\text{R}}$, \textit{ i.e.}, the pathway to attain that equilibrium structure at a given osmotic pressure. Although the pathways and symmetry of the preferred equilibrium ordering differ depending on particle geometry, these inferences are universal and equally applicable to other dense systems of hard-interacting anisotropic particles.

\section*{Conclusions}

We show using experimental studies that in a dense assembly of hard-interacting squares, the range of accessible orientational states regulates their reorganization dynamics and thus the crystallization pathways to optimal equilibrium symmetry structures. Specifically, $\Delta\theta_{\text{R}}$ governs whether the system crystallizes into RB or HX symmetry at a given osmotic pressure, where the local density, which sets the upper limit on the value of $\Delta\theta_{\text{R}}$, has only an indirect effect on the process, as corroborated by our Brownian dynamics simulations and free energy calculations. Furthermore, to demonstrate the generalizability of the results, we show that our findings are also applicable to other anisotropic particles, such as Brownian triangles \cite{Zhao2012a}.
	
Phase transitions in systems of isotropic hard-interacting thermalized particles, commonly known as hard spheres (hard disks in 2D), are governed solely by the density (area fraction) that controls their accessible translational states. In this study, we establish that, in the case of anisotropic hard-interacting particles, the range of accessible orientational states principally governs the pathways of crystallization, and thus regulates the phase transitions, playing a similar role as density in hard sphere systems. In general, when the self-assembly of anisotropic constituents equilibrates, a wider range of accessible orientational states of the particles leads to a higher-order hexagonal symmetry by maximizing the orientational entropy, whereas for an enhanced degree of asymmetry in the shape, such as in particles with sharper corners, orientational dynamics become more restricted or time-consuming, thus inducing side-to-side alignment. Consequently, longer collective positional fluctuations facilitate the maximization of configuration entropy through its translational component, leading to shape-symmetry-dependent crystalline ordering akin to the nematic transition of Onsager's rods \cite{Onsager1949,Frenkel1987,Frenkel2014}. Notably, the kinetic arrest of rotational dynamics may also lead to an orientationally disordered jammed state for certain anisotropic particles, thus impeding translational dynamics and consequently inducing a glass transition in these systems \cite{Zhao2009, Zheng2011}. Thus, our findings are generally applicable to predict the crystallization pathways of other anisotropic hard-interacting particles in 2D and can be directly adapted to 3D. Interestingly, the emergence of translational ordering from orientational disorder, as observed in the purely entropy-driven crystallization of Brownian squares into HX symmetry, has also been observed in a system of interacting anisotropic nanoparticles \cite{Ou2019}. Although the interaction among the square particles in this system is solely entropic, if the constituents have attractive or repulsive interactions alongside excluded volume effects, the competition between enthalpy and entropy, which can be estimated based on our predictions, dictates the crystallization pathways.

Studies on the phase behavior of dense systems of anisotropic particles began with Onsager’s spherocylinders \cite{Onsager1949}. Although exotic equilibrium phase behaviors in these entropic systems have been studied extensively in simulations and experiments with various shapes \cite{Onsager1949, Frenkel1988, WOJCIECHOWSKI2004, MartinezRaton2005, John2008, Zhao2009, Zhao2011, Agarwal2011, Damasceno2011, Zhao2012, Zhao2012a, Damasceno2012, Wang2015, Klotsa2018, Shen2019}, the underlying equilibration dynamics leading to crystalline and liquid-crystalline phases with varied symmetries remain a longstanding pursuit. Our results address this quest and provide a comprehensive understanding of the pathways to equilibrium phase behaviors that are solely governed by access to a wider range of orientational states of anisotropic particles. Thereby, our findings explain how a slightly higher degree of corner rounding alters the equilibrium phase behavior \cite{Zhao2011,Avendano2012}, and the coexistence of RX and RB crystallites in the same neighborhood in the CE phase with local variations in the range of accessible orientational diffusion of the Brownian squares \cite{Zhao2011}--phenomena that are generalizable to other anisotropic particles.

A detailed estimation of the particle separation- and shape-dependent frictional effects and complex hydrodynamic interactions in fluctuating 2D systems of platelet particles by calculating  the grand resistance and mobility tensors, potentially employing simulations, and a detailed examination of the validity of our conclusions for other complex anisotropic shapes, e.g., concave particles, remain interesting future directions. This study opens avenues for exploring the effect of directional interactions among constituents on the crystallization pathways leading to specific equilibrium orderings and the structural and mechanical properties of these crystalline phases, such as their response to defects in two and three dimensions.

\section*{Materials and methods}
\subsection*{Sample preparation and data collection}
The square platelet particles were made of cross-linked SU-8 polymer photoresist and fabricated using top-down photolithography \cite{Zhao2011,Wang2015}. The side length and thickness of the square platelets were measured to be \SI{2.5 \pm 0.1}{\micro \meter} and \SI{1.0 \pm 0.1}{\micro \meter}, respectively, according to SEM (Fig. \ref{fig:System} (E)). The squares have slightly rounded corners owing to the limited feature size of the stepper (\SI{\approx  300}{\nano \meter}).

A mixture of a dilute suspension of squares and a depletion agent (polystyrene spheres, diameter \SI{\approx  42}{\nano\meter}, concentration \SI{\approx 0.55} \% wt/vol) was filled into a thin-walled narrow rectangular glass cuvette. The diameter and concentration of the depletion agent were optimized to retain the square particles in a 2D layer on the bottom surface \cite{Zhao2011}. The squares can freely move in the plane to form a quasi-2D system because the adsorbed anionic dodecyl sulfate surface charges prevent them from pinning onto the negatively charged glass surface. Because the edges of the particles are rougher than their faces, there is no in-plane depletion attraction among the squares; hence, they interact solely through an excluded volume field. The cuvette was then slightly tilted about its long axis (\SI{\approx 3}{\degree}), which created a surface density gradient owing to gravity as the 2D assembly of square particles was osmotically compressed (Fig. S7). Under quasi-static compression for a few (8-10) weeks, the system slowly equilibrated into different phases over a gradually varying gradient of osmotic pressure \cite{Zhao2011}. The positional and orientational fluctuations of the particles in both RB and HX were imaged through a 60X (Nikon plan apochromat) objective and recorded over a field of view of dimension $\mathrm{117 \times 88} \mathrm{\; \mu m}$ (or $\mathrm{1600 \times 1200}$ pixel) at 15 fps using a CMOS camera (Flea Cam, Point Grey) attached to a Nikon inverted microscope. Multiple image sequences of the RB and HX regions were captured from different cuvettes.

\subsection*{Image and data analysis}
Typical image sequences that were verified to be free from any drift, following local collective motion analysis \cite{Khan2016}, were further processed with thresholding and watershed segmentation using Python and Fiji for particle detection (SM, Fig. S1). The particle positions ($\vec{r}_j (t)$, $j$ denotes the particle tag) were tracked using an existing tracking algorithm (TrackPy) \cite{Crocker1996}, whereas the orientations ($\theta_j (t)$) were tracked using in-house Python code. 

For other quantitative analyses and visualizations, such as Voronoi construction, Delaunay triangulation analyses, calculation of positional and bond-orientational order parameters, and plotting, we used the scientific computing toolkit of Python (Scipy, Numpy, Matplotlib, Pandas, etc.) and OriginLab.

\subsection*{Simulations}
Square particles with rounded corners were realized using the equation $\left| x \right|^{p} + \left| y \right|^{p} = (L/2)^{p} $, where $p$ and $L$ denote the corner rounding and side length, respectively. We used $p = 4.84$ and $L = 2.50$-length units (\SI{1} {\micro \meter}) to match the shape of the particles used in the experiments. The self-assemblies of 36 such particles were simulated, where the inner 16 particles were allowed to move while the remaining 20 boundary particles remained static, preserving the area fractions ($\Phi$) and lattice angles ($\alpha$) corresponding to the RB and HX phases. At every time step ($dt$ = \SI{2.5E-3}{s}), the mobile particles were subjected to random positional and orientational displacements following zero-mean normal distributions with a standard deviation $\sqrt{2 D_{\text{t,r}} dt}$, where $D_{\text{t,r}}$ represents the corresponding diffusion coefficient. In addition, the orientational fluctuations of the mobile particles were restricted within a predefined range set by the corresponding $\Delta\theta_{\text{R}}$ (Fig. \ref{fig:Dynamics} (E, F)). The updated configuration after each time step was accepted only if there was no overlap among the neighboring particles and the range of orientational fluctuations of each particle did not exceed the set value, $\Delta\theta_{\text{R}}$ (Fig. S5). Square particles with the same length but enhanced corner rounding as per $p = 4.50$ were used to simulate the equilibration of an initial disordered RB configuration with a wider $\Delta\theta_{\text{R}}$ corresponding to HX-C (Fig. \ref{fig:WiderRange}).  The systems were considered to reach equilibrium after $\approx 7 \times 10^5$ steps ($\equiv$ \SI{1750}{s} in real time) on average when the positional distributions of the particles became Gaussians, as observed in the experiments, and no longer evolved with time, \textit{i.e.}, with additional simulation steps. 

The final equilibrium configurations are defined by the time-averaged particle positions from the last 4500 steps ($\equiv$ 11.25 s in real time) and are shown as schematics rendered in Blender. Particles in the intermediate structural configurations are identified as crystalline if their local $m$-fold bond orientational ($\psi_m$) and positional ($\zeta_m$) order parameters, calculated from their positions averaged over 4500 frames, satisfy a set threshold criterion. The order parameter criteria for RB and HX are defined as $\left| \psi_4\right| \geq$ 0.67, $\operatorname{Re}(\zeta_4) \geq$ 0.87, and $\left| \psi_6\right| \geq$ 0.92, $\operatorname{Re}(\zeta_6) \geq$ 0.89.

\section*{Data availability}
All data required to evaluate the conclusions of this study are presented in the manuscript or in the Supplementary Materials.

\section*{Acknowledgements}

MK acknowledges funding from IIT Kanpur through an initiation grant (IITK-PHY-2017081) and financial support from UCLA during the course of experiments. TGM thanks UCLA for financial support. DC acknowledges the financial support through the PMRF scheme. The authors thank P.-Y. Wang for assistance in fabricating the lithographic particles at UCLA and Tamoghna Das for fruitful discussions.

\section*{CRediT authorship contribution statement}
\textbf{Debojit Chanda:} Conceptualization, Data curation, Formal analysis, Investigation, Methodology, Validation, Visualization, Writing – original draft, Writing – review and editing. \textbf{Thomas G. Mason:} Conceptualization, Funding acquisition, Methodology, Project administration, Resources, Supervision, Writing – review and editing. \textbf{Manas Khan:} Conceptualization, Data curation, Funding acquisition, Investigation, Methodology, Project administration, Resources, Supervision, Validation, Writing – original draft, Writing – review and editing.

\section*{Competing interests}
The authors declare no competing interests.


\bibliographystyle{elsarticle-num} 
\bibliography{RouteToCrystallization.bib}

\end{document}


\begin{frontmatter}
	
	
	\title{{\normalsize \bfseries Supplementary Materials} \\ Entropic crystallization of Brownian squares through pathways governed by orientational dynamics}
	
	\author[a]{Debojit Chanda}
	\author[b,c]{Thomas G. Mason}
	\author[a]{Manas Khan\corref{*}} 
	
	\affiliation[a]{organization={Department of Physics, Indian Institute of Technology Kanpur}, city={Kanpur}, postcode={208016}, country={India}}
	\affiliation[b]{organization={Department of Physics and Astronomy, University of California}, city={Los Angeles}, state={CA}, postcode={90095}, country={USA}}
	\affiliation[c]{organization={Department of Chemistry and Biochemistry, University of California}, city={Los Angeles}, state={CA}, postcode={90095}, country={USA}}
	\cortext[*]{mkhan@iitk.ac.in}

\end{frontmatter}

	\section{Image analysis}
	\label{sec:ImageAnalysis}

	Image sequences of osmotically compressed self-assembly of square particles were captured for $\approx$160 s at 15 frames per second, accumulating $\approx$ 2400 image frames in a sequence, which is adequate for studying the slow configurational dynamics of square particles at high densities. The size of each frame is 1600 $\times$ 1200 pixels, with one pixel corresponding to $\SI{0.073}{\micro \meter}$. Extraction of relevant data from the captured image sequences for further analyses were performed in two steps: first, identifying the particles, and then, tracking the translational and rotational dynamics of each particle.
	
	\subsection{Particle detection}
	
	First, thresholding was applied to the hexagonal (HX) image sequence in Fiji (formerly ImageJ) \cite{Schindelin2012} to separate particles from the background. When the particles are closely spaced and mostly side-aligned, thresholding in Fiji does not recognize all particles separately. Hence, Python (using package \verb|scikit-image|) \cite{VanderWalt2014} was used for thresholding the rhombic (RB) image frames. In addition, watershed segmentation was applied to densely packed regions to enhance the reliability of particle detection. These transformed pseudo-colored monochrome images into binary frames (Fig. \ref{ImageProcessing} (A \& B)). The white areas shown in Fig. \ref{ImageProcessing}, were detected as particles and marked with unique integer tags, to later track the dynamics of each particle. A few particles $(< 0.3\%)$ remain undetected in some frames. We consistently detected and tracked 1305 and 1268 particles in RB and HX image frames, respectively.

	\subsection{Position and orientation tracking}
	
\begin{figure}[!ht]
		\includegraphics[width=\linewidth]{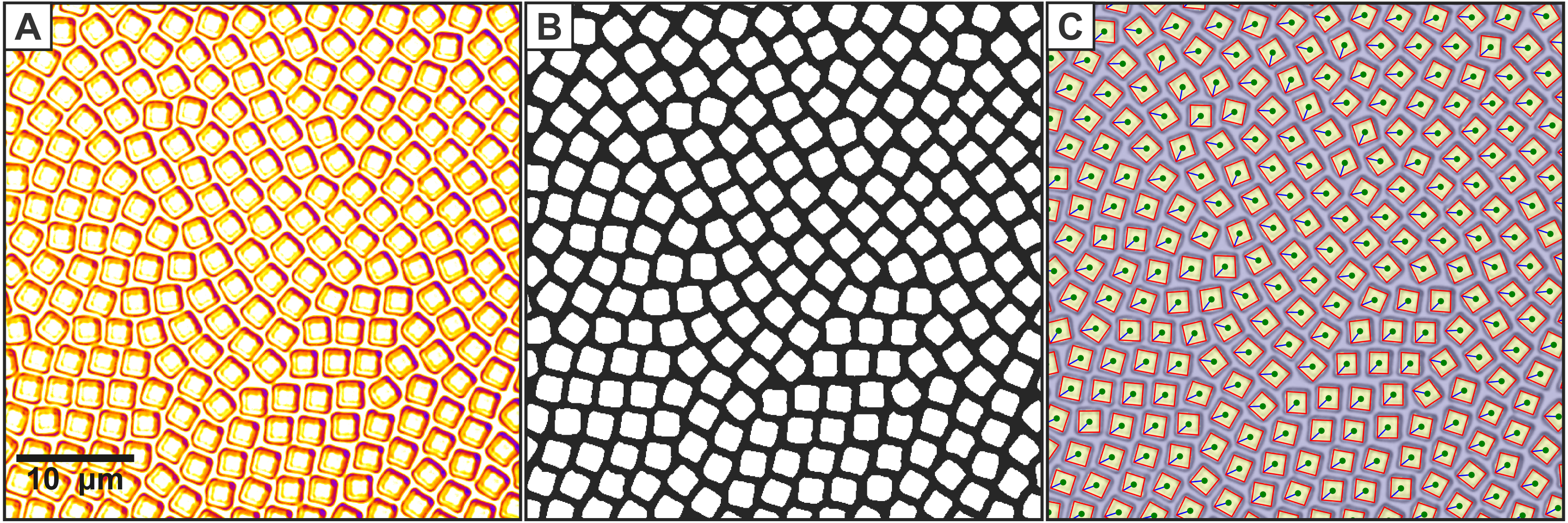}
		\caption{Detection and identification of the center and orientation of each particle in a frame. A pseudo-colored monochrome image frame (A) is converted to binary (B) by thresholding to differentiate the particles from the background. (C) The center of each particle (green dot) is identified as the center of the fitted parallelogram (red line). The orientation of a particle is defined as the angle between a chosen half-diagonal (blue line) and positive horizontal direction.}
		\label{ImageProcessing}
	\end{figure}

	An in-house Python script was used to fit a parallelogram to each detected and tagged particle in all the frames, as shown in Fig. \ref{ImageProcessing} (C). The centroids of the fitted parallelograms defined the centers of the particles, which were tracked using an available particle tracking algorithm \cite{Crocker1996} implemented in the \verb|Trackpy| package \cite{Allan2023a}. This provided the positional time-series for all particles,\textit{ i.e.}, $\vec{r}_j (t)$, where $j (= [1,N])$ denotes the particle tag. A specific corner of each parallelogram in the first frame of the image sequence was chosen to define the orientation of the corresponding particle as the angle $(\theta)$ made by the line, which connects the centroid and the chosen corner, with the positive horizontal direction (Fig. \ref{ImageProcessing} (C)). Subsequently, the orientations of all particles were tracked through all image frames in a sequence using an in-house Python code to obtain the orientational time-series ($\theta_j (t)$).

\section{Identifying order and symmetry \label{sec:order}}

\begin{figure}[!h]
	\centering
	\includegraphics[width=0.85\linewidth]{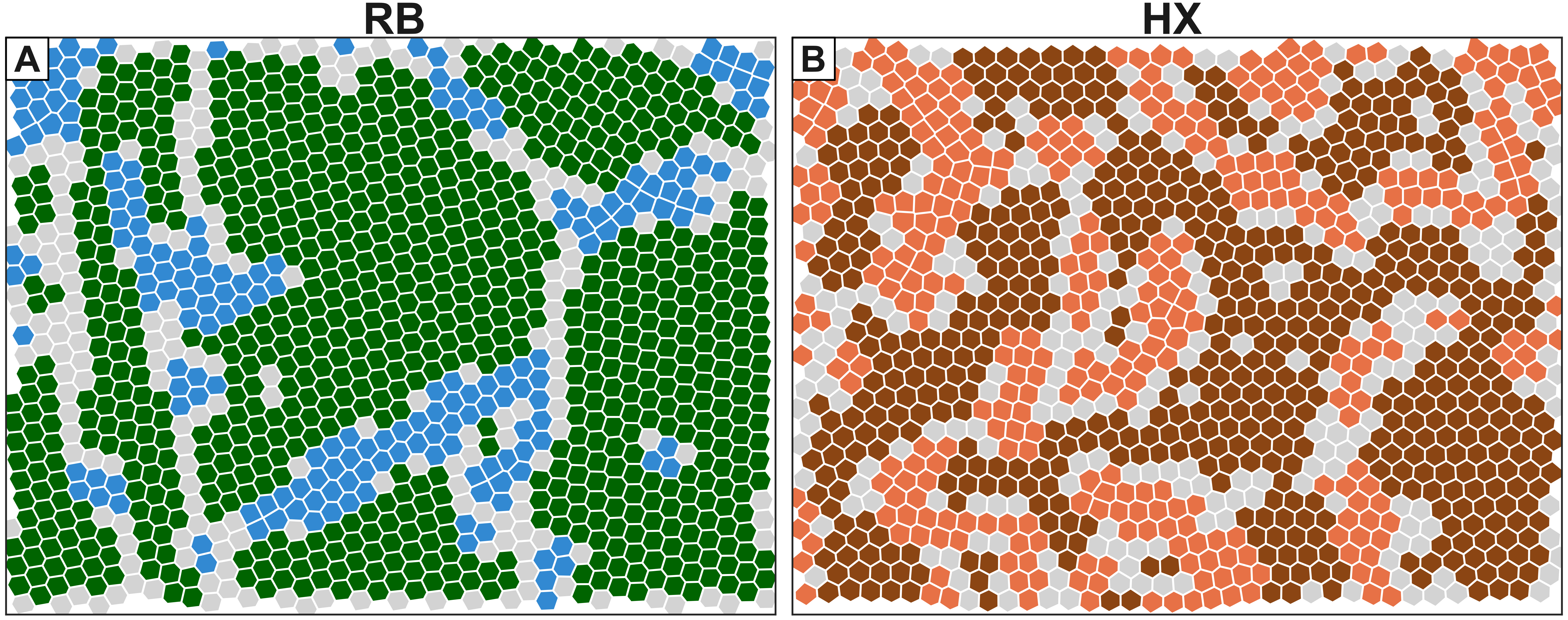}
	\caption{Voronoi diagrams for RB (A) and HX (B) phases. The Voronoi cells in RB-C and HX-C crystallites are filled with green and brown, whereas blue and orange-colored Voronoi cells represent the NC domains of RB and HX, respectively. The grey-colored cells do not belong to either the C or the NC domains.}
	\label{Voronoi}
\end{figure}

\begin{figure}[!ht]
	\includegraphics[width=\linewidth]{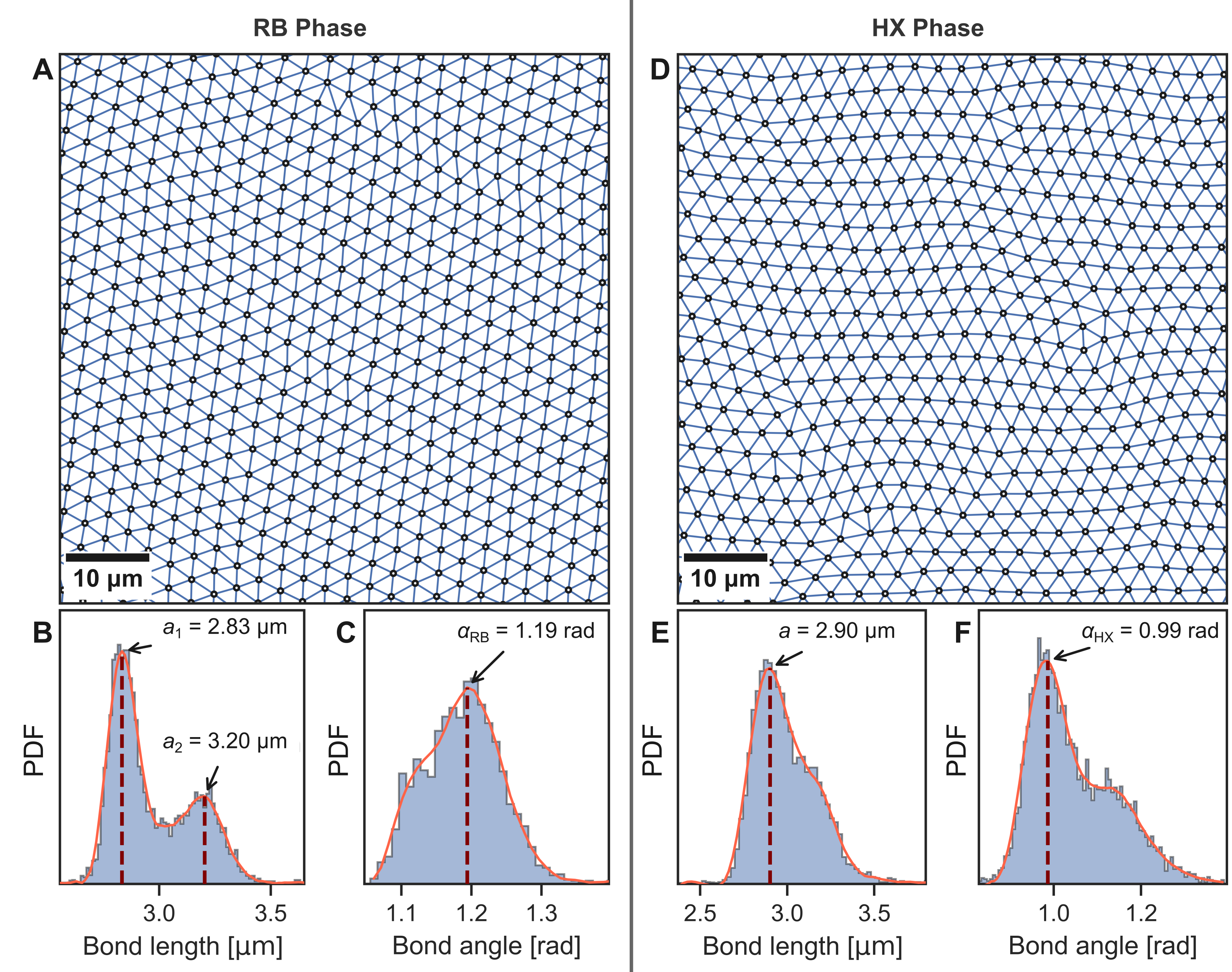}
	\caption{Delaunay triangulation, probability distributions of bond lengths, and bond angles for RB (A, B \& C) and HX (D, E \& F). Delaunay triangulations on time-averaged particle positions (A \& D) provide all bond lengths and bond angles. Their probability distributions are plotted with blue bars, and the smoothed plots are shown with orange curves. (B, E) The most probable bond lengths, $a_1$ \& $a_2$ for RB, and $a$ for HX, were obtained from the positions of the peaks of the corresponding smoothed probability distributions (orange curves). (C, F) Similarly, the peaks of the smoothed probability distributions provide the most probable bond angles $\alpha_\text{RB}$ and $\alpha_\text{HX}$, where $\alpha_\text{RB}$ denotes the smaller among the two lattice angles of RB unit cell.}
	\label{BondParameters}
\end{figure}

For all analyses to identify the order and symmetry, we considered the particle positions averaged over all frames in the corresponding image sequence. First, we obtained the Voronoi diagrams (Fig. \ref{Voronoi}) and Delaunay triangulation (Fig. \ref{BondParameters} (A \& D)) for both the RB and HX phases using the Python package \verb|scipy|. While the FFT of the corresponding image frames (Fig. 1 (A \& B Insets)) makes the prevailing order apparent, the shapes of the Voronoi cells and periodicity in Delaunay triangulation provide a qualitative description of the underlying ordering to distinguish the crystallites (C) from the non-crystalline (NC) domains. 

Delaunay triangulations further provided all bond lengths and bond angles that were used to compute the order parameters. The bond lengths were obtained as two shorter side lengths of the triangulation simplices, and the angle between them provided the bond angles. The positions of the peaks in the probability distribution of these bond lengths, $a_1 = \SI{2.83} {\micro \meter}$ and $a_2= \SI{3.20} {\micro \meter}$ for RB, and $a = \SI{2.90} {\micro \meter}$ for HX, represent the corresponding most probable values, as shown in Fig. \ref{BondParameters} (B \& E). Similarly, we obtained the most probable bond angles, $\alpha_\text{RB} = \SI{1.19} {rad}$ and $\alpha_\text{HX} = \SI{0.99} {rad}$, from their probability distributions (Fig. \ref{BondParameters} (C \& F)). The order parameters were calculated based on standard definitions \cite{Murray1987,Strandburg1988,Zhao2011}.

\subsection{Local bond-orientational order parameter}

For the $j$-th particle, the local $m$-fold bond-orientational order parameter $\psi_{m}^j$ is given by a complex number, and defined as,

\begin{equation}
	\psi_{m}^{j} = \frac{1}{N_{j}} \sum_{k=1}^{N_{j}} e^{i m \theta_{jk}}
	\label{eq1}
\end{equation}

where $N_{j}$ is the number of neighbors of the $j$-th particle. $\theta_{jk}$ is the angle between an arbitrary fixed axis (here, the $x$-axis) and the bond between the $j$-th particle and its $k$-th neighbor, \textit{ i.e.}, the line connecting the centers of the two particles. We use $\left| \psi_{m}^{j}\right|$ as the relevant order parameter for this system.

To calculate 4-fold order parameters we considered only the four closest neighboring particles, \textit{i.e.}, $N_{j} = 4$. For calculating 6-fold order parameters in the HX phase, the number of neighbors was determined using the Voronoi construction.

\subsection{Local positional order parameter}
The $m$-fold positional order parameter for the $j$-th particle, $\zeta_{m}^{j}$, is defined as,

\begin{equation}
	\zeta_{m}^{j} = \frac{1}{N_{j}} \sum_{k=1}^{N_{j}} \exp\left( i \vec{G}_m \cdot (\vec{r}_j - \vec{r}_k)\right)
\end{equation}
where $\vec{r}_j$ denotes the position of the $j$-th particle with $N_{j}$ number of neighbors and $\vec{G}_m$ is the reciprocal lattice vector. Because only the real part of $\zeta_{m}^{j}$ is relevant here, we used $\mathrm{Re} (\zeta_{m}^{j})$ as the local positional order parameter for our system.

While to calculate 6-fold order parameter we counted $N_{j}$ from the Voronoi construction, in case of 4-fold order parameter, we considered 4 nearest-neighbors and 4 next-nearest-neighbors, \textit{i.e.}, $N_{j}$ = 8 in total, to compute the local positional order parameter. Moreover, $\vec{G}_6$ for HX was calculated from a perfect hexagonal lattice with bond length $a$, whereas a primitive lattice was constructed using the bond angles and the most probable bond length to obtain $\vec{G}_4$.

\subsection{Local order parameter threshold to distinguish ordering}

The $m$-fold local order parameter of a particle is a quantitative measure of how closely the ordering of its neighbors matches a periodic structure with $m$-fold symmetry. Thus, a higher value of the $m$-fold local order parameter indicates that the particle belongs to a crystalline domain with $m$-fold symmetry. In contrast, if the $m$-fold local order parameter is smaller, then the particle and its neighbors do not form an $m$-fold order. To identify the particles that constitute RB or HX crystallites, we used their 4-fold or 6-fold local order parameters, respectively, as distinguishing criteria. If the 4-fold bond-orientational and positional order parameters of a particle were greater than the specifically chosen threshold values, it was identified as forming a local rhombic crystalline (RB-C) structure with its neighbors. Similarly, particles in HX were recognized to form a local hexagonal crystalline structure (HX-C) when both their 6-fold order parameters were greater than the corresponding threshold values. The threshold values given in Table \ref{LocalOPThreshold} were defined such that they were consistent with the qualitative description of ordering, as seen in the Voronoi diagrams (Fig. \ref{Voronoi}) and Delaunay triangulations (Fig. \ref{BondParameters}). The particles that did not satisfy both the local order parameter threshold criteria were considered to be in the non-crystalline domains (RB-NC and HX-NC), which would eventually crystallize later. Notably, $\psi_4$ values for the particles in the RB phase are considerably smaller on average because the bond orientations in the RB phase deviate from that in a perfect lattice with 4-fold symmetry, \textit{i.e.}, a square crystal. However, with a smaller average value of $\psi_4$, an appropriately chosen smaller threshold value provided a good identifier of the RB-C particles from those constituting RB-NC.

Finally, we considered only those particles whose both local order parameters were larger than the corresponding threshold values and had more than three similar neighbors to constitute the crystallites. In contrast, all particles that did not satisfy the threshold criteria and had at least three similar neighbors were recognized to form NC domains. These additional qualification criteria regarding the neighbors were imposed to ensure that the configurational dynamics of a crystalline or non-crystalline particle were not influenced by the majority of its neighbors with different structures and, hence, dissimilar dynamical properties. Furthermore, these conditions helped remove small islands, resulting in smooth boundaries in the C and NC regions. Particles that did not satisfy these criteria were not considered in further analysis (Fig. 1, Fig. \ref{Voronoi}). Thereby, we have 782 and 561 particles forming the crystallites, and 264 and 415 particles constituting the NC regions, in RB and HX, respectively.

\begin{table}[!h]
	\centering
	
	\tabulinesep=1.5mm
	\begin{tabu}{l|c|c}
		\text{Local $m$-fold order parameters: }& $|\psi_{m}^{j}|$ & $\mathrm{Re} (\zeta_{m}^{j})$ \\ \hline
		\text{Rhombic crystalline order (RB-C), $m=4$}   & 0.61            & 0.82                         \\ \hline
		\text{Hexagonal crystalline order (HX-C), $m=6$} & 0.90            & 0.90                         \\ \hline
	\end{tabu}
		\caption{\footnotesize Threshold values of the local $m$-fold bond-orientational ($|\psi_{m}^j|$) and positional $\mathrm{Re} (\zeta_{m}^j)$ order parameters to identify particles that constitute rhombic and hexagonal crystallites.}
	
	\label{LocalOPThreshold}
\end{table}

\subsection{Global order parameters}

The $m$-fold global order parameters, obtained by appropriately averaging the corresponding local order parameter values over a larger region, provide a quantitative measure of the ordering and symmetry of self-assemblies. We defined the $m$-fold global bond-orientational and positional order parameters for our system as follows:

\noindent \textbf{Global $m$-fold bond-orientational order parameter ($\Psi_m$):}
\begin{equation}
	\Psi_m = \frac{1}{N} \sum_{j=1}^N |\psi_{m}^{j}|
\end{equation}

\noindent \textbf{Global $m$-fold positional order parameter ($\mathrm{Z}_m$):}
\begin{equation}
	\mathrm{Z}_m = \frac{1}{N} \sum_{j=1}^N \mathrm{Re}(\zeta_{m}^{j})
\end{equation}
where $N$ is the number of particles in the system and $\psi_{m}^{j}$ and $\zeta_{m}^{j}$ are the local $m$-fold bond-orientational and positional order parameters of the $j$-th particle, respectively. We averaged over $|\psi_{m}^{j}|$ and $\mathrm{Re}(\zeta_{m}^{j})$ because the relative orientations of the unit cells and crystallites were not relevant here.

\subsection{Global order parameters to identify symmetry}

While the FFT (Fig. 1 (A \& B insets)), Voronoi diagrams (Fig. \ref{Voronoi}), and Delaunay triangulations (Fig. \ref{BondParameters} (A \& D)) of time-averaged particle positions in the corresponding image frames helped us identify prevailing symmetries, and thus recognize RB and HX phases, we further calculated the global $m$-fold order parameters as quantitative identification criteria. The values of the 4-fold global order parameters for the RB phase, and both the 4-fold and 6-fold global order parameters for HX are given in Table \ref{GlobalOP_RB_HX}. Higher values of 4-fold global order parameters verified underlying structural ordering with 4-fold symmetry in RB, whereas lower values of 4-fold global order parameters and higher values of 6-fold counterparts validated the prevailing hexagonal ordering in HX.

\begin{table}[!h]
	\centering
	\tabulinesep=1.5mm
	\begin{tabu}{l|c|c|c|c}
		\text{Global $m$-fold order parameters:}                                               & $\Psi_4$    & $\mathrm{Z}_4$ & $\Psi_6$    & $\mathrm{Z}_6$ \\ \hline
		\text{Rhombic Phase (RB)}   & $\bm{0.63  \pm 0.01}$ & $\bm{0.88  \pm 0.01}$    & -           & -              \\ \hline
		\text{Hexagonal Phase (HX)} & $0.54  \pm 0.01$        & $0.77  \pm 0.01$           & $\bm{0.90}$ & $\bm{0.89}$    \\ \hline
	\end{tabu}
	\caption{\footnotesize The values of the 4-fold global order parameters for the RB phase, and both the 4-fold and 6-fold global order parameters for HX are shown with standard error estimations. Standard errors that are less than $0.005$, \textit{i.e.}, zero when rounded to two decimal places, are not shown. The order parameter values that define the prevailing symmetry structures are indicated in bold font.}
	\label{GlobalOP_RB_HX}
\end{table}

We computed the global order parameter values for the C and NC domains to corroborate their structural symmetry and lack of symmetry, respectively. The values of the 4-fold and 6-fold global order parameters are listed in Table \ref{GlobalOP_C_NC}. Higher and lower 4-fold global order parameters confirmed the presence and absence of 4-fold structural symmetry in RB-C and RB-NC, respectively. Similarly, the identification of HX-C and HX-NC was corroborated by their respective higher and lower 6-fold global order parameter values. Additionally, a lower value of the global 6-fold bond orientational order ($\Psi_6$) in both RB-NC and HX-NC verified the absence of the hexatic phase in the system.

\begin{table}[!h]
	\centering
	\tabulinesep=1.5mm
	\begin{tabu}{l|c|c|c|c}
		\text{Global $m$-fold order parameters:} & $\Psi_4$ & $\text{Z}_4$ & $\Psi_6$ & $\text{Z}_6$ \\ \hline
		\text{Rhombic crystallites (RB-C)}        & $\bm{0.73}$     & $\bm{0.97}$           & -        & -              \\ \hline
		\text{Rhombic non-crystalline domains (RB-NC)}   & $0.41 \pm 0.01$     & $0.67 \pm 0.01$           & $0.78 \pm 0.02$     & $0.80 \pm 0.01$           \\ \hline
		\text{Hexagonal crystallites (HX-C)}      & $0.54 \pm 0.01$     & $0.82 \pm 0.01$           & $\bm{0.95}$     & $\bm{0.95}$           \\ \hline
		\text{Hexagonal non-crystalline domains (HX-NC)} & $0.56 \pm 0.01$     & $0.70 \pm 0.01$           & $0.80 \pm 0.01$     & $0.80 \pm 0.01$           \\ \hline
	\end{tabu}
	\caption{\footnotesize The values of global order parameters for the C and NC domains in RB and HX. The order parameter values are shown with corresponding standard errors that are greater than $0.005$, \textit{i.e.}, non-zero when rounded to two decimal places. Order parameter values that indicate the presence of structural symmetry are shown in bold.}
	\label{GlobalOP_C_NC}
\end{table}

\section{Cage-relative mean square displacement}

Cage-relative mean square displacements (CR-MSD) reveal the dynamics of a particle with respect to a cage formed by its neighbors, and thus, can eliminate collective dynamics. The cage-relative position ($\delta\vec{r}_j(t)$) and orientation ($\delta\theta_j(t)$) at time $t$, and the corresponding displacements ($\Delta(\delta\vec{r}_j) (\Delta t)$ \& $\Delta(\delta\theta_j) (\Delta t)$) of the $j$-th particle in time interval $\Delta t$ are defined as \cite{Mazoyer2009,Illing2017,Kelleher2017},
\begin{align}
	\delta\vec{r}_j(t) &= \vec{r}_j (t) - \frac{1}{N_{j}} \sum_{k=1}^{N_j} \vec{r}_k(t), &  \Delta(\delta\vec{r}_j)(\Delta t) = \Delta \vec{r}_j (\Delta t) - \frac{1}{N_{j}} \sum_{k=1}^{N_j} \Delta \vec{r}_k (\Delta t) \\
	\delta\theta_j(t) &= \theta_j (t) - \frac{1}{N_{j}} \sum_{k=1}^{N_j} \theta_k(t), &  \Delta(\delta\theta_j)(\Delta t) = \Delta \theta_j (\Delta t) - \frac{1}{N_{j}} \sum_{k=1}^{N_j} \Delta \theta_k (\Delta t)
\end{align}
The position, orientation, and corresponding displacements of the cage are calculated as the mean of the respective variables over $N_j$ number of neighbors with index $k$. Therefore, we define the time- and ensemble-averaged translational and orientational CR-MSDs as $\langle (\Delta (\delta\vec{r}))^2 \rangle_{t,j} (\tau)$ and $\langle (\Delta (\delta\theta))^2 \rangle_{t,j} (\tau)$, respectively.

We obtained the number of neighbors for the $j$-th particle, $N_j$, and their positions ($\vec{r}_k(t)$) and orientations ($\theta_k(t)$) from Delaunay triangulation. The CR-MSDs were averaged over both time ($t$) and particles ($j$), considering the system to be stationary over the observation time because of the slow dynamics of the particles. The time- and ensemble-averaged CR-MSDs obtained from the experiments and simulations are shown in Fig. 2 and Fig. \ref{CRMSD_Sim}, respectively.

\section{Standard errors in the MSDs}

We estimated the standard error (SE) of the translational and orientational MSDs, \textit{i.e.,} $\langle (\Delta \vec{r} (\tau) )^2\rangle = \langle (\vec{r}_j (t+\tau)  - \vec{r}_j (t) )^2 \rangle_{j, t}$ and $\langle (\Delta \theta (\tau))^2 \rangle = \langle (\theta_j (t+\tau)  - \theta_j (t))^2 \rangle_{j, t}$, respectively, for a given time lag $\tau$ as,
\begin{equation*}
	\langle (\Delta \vec{r} (\tau) )^2\rangle_{\mathrm{SE}} = \dfrac{\sigma\left(\Delta \vec{r}(\tau)^2\right)}{\sqrt{N \times \frac{T-\tau}{dt}}}, \quad \text{and} \quad
	\langle (\Delta \theta (\tau))^2 \rangle_{\mathrm{SE}} = \dfrac{\sigma\left(\Delta \theta (\tau)^2\right)}{\sqrt{N \times \frac{T-\tau}{dt}}},
\end{equation*}
where, $N$ is the total number of particles with the same structural ordering, $T$ is the total duration of the trajectory, and $dt$ is the temporal resolution. The numerators denote the standard deviations of the squared translational and orientational displacements, respectively, and the denominator corresponds to the square root of the total number of statistically independent realizations contributing to the average, for all $N$ particles. A similar formulation, with squared cage-relative displacements in place of squared displacements, has been employed to compute the SE of the cage-relative MSDs (CR-MSDs).

The standard errors in the translational and orientational MSDs and CR-MSDs were evaluated at the largest time lag considered ($\tau = \SI{100}{\second}$), where the available statistics for averaging is the lowest. Consequently, the SE at shorter time lags are expected to be smaller. At $\tau = \SI{100}{\second}$, the standard errors in the orientational MSDs for the RB-NC, HX-C, and HX-NC were found to be \SI{0.001}{\radian\squared}. The corresponding errors in the orientational CR-MSDs were \SI{0.001}{\radian\squared}, \SI{0.001}{\radian\squared}, and \SI{0.002}{\radian\squared}, respectively. For all other translational and orientational MSDs and CR-MSDs, the standard errors were zero up to three decimal places.

\section{Range of translational \& orientational fluctuations of particles}

\begin{figure}[!h]
	\centering
	\includegraphics[width=0.98\linewidth]{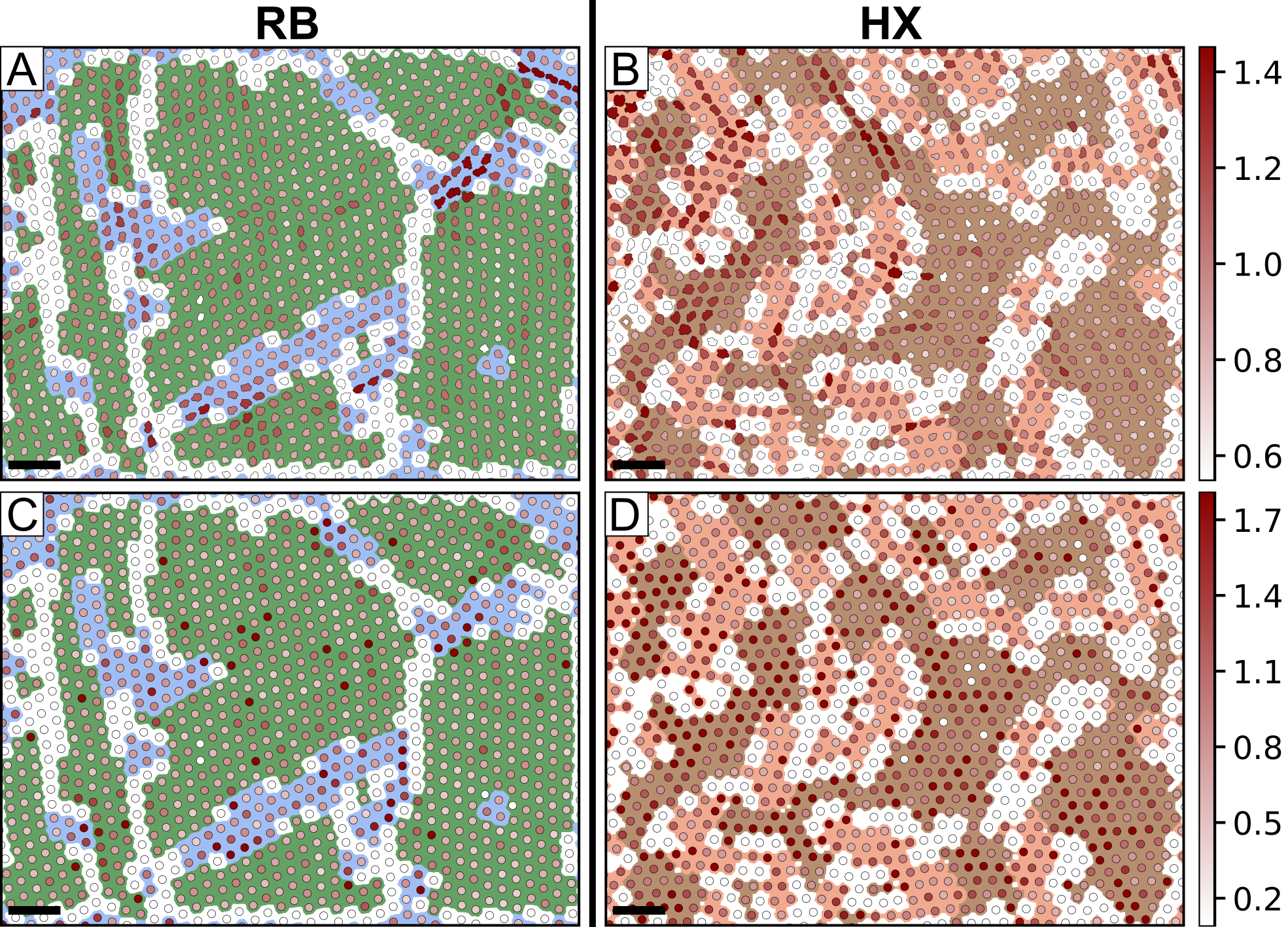}
	\caption{Range of translational and orientational fluctuations of the particles constituting the C and NC regions in RB and HX. The top row shows the span of the trajectories,\textit{ i.e.}, the range of translational motion of the particles in (A) RB and (B) HX. The colors correspond to the top color bar, where the unit is \unit{\micro \meter}. The range of accessible orientational fluctuations of the particles is exhibited by color-coded filled circles placed at the time-averaged positions of the particles in (C) RB and (D) HX. The colors represent the range of available rotational fluctuations in rad as per the bottom color bar. Particles constituting C and NC regions are overlaid on green and blue backgrounds for RB (A, C), whereas brown and orange backgrounds indicate C and NC regions in HX (B, D). All scale bars denote 10 \unit{\micro\meter}.}
	\label{DynamicalRange}
\end{figure}

Ensemble-averaged mean-square displacements and their cage-relative versions (Fig. 2 (A - D)) represent the average dynamical behavior of all particles constituting a structural domain, such as, the C or NC regions in RB and HX. Here, we show the accessible range of translational and orientational fluctuations of each particle, overlaying them on a color-coded background, distinguishing the C and NC regions in the RB and HX (Fig. \ref{DynamicalRange}).

\section{BD simulation of crystallization with varied range of accessible orientational states}

We employed two-dimensional Brownian dynamics simulations to understand how hard-interacting square particles crystallize into different equilibrium structures depending on the range of their accessible orientational states ($\Delta \theta_{\text{R}}$). To simulate a square particle with rounded corners, we used the equation $|x|^p + |y|^p = (L/2)^p$, where the parameter $p (\geq 2)$ defines the degree of corner rounding. For the limiting case $p=2$, the equation represents a circle and $p \rightarrow \infty$ corresponds to a perfect square with sharp corners. We used $L = 2.5$ \unit{\micro\meter} and $p = 4.84$, which replicated the shape of the square particles used in the experiments (Fig. 4, Inset). The simulations were performed with 36 such corner-rounded particles (Fig. 3 and 4). Among these, inner 16 particles were allowed to move, while the remaining 20 boundary particles were kept stationary, thus fixing the area fraction ($\Phi$) and the lattice angle ($\alpha$) of the system throughout the simulation. The boundary particles were placed to set the values of $\Phi$ similar to those observed in our experiments, whereas they were oriented at $\alpha/2$ (with respect to the $x$- axis), where $\alpha$ is the lattice angle.

\begin{figure}[!h]
	\centering
	\includegraphics[width=0.6\linewidth]{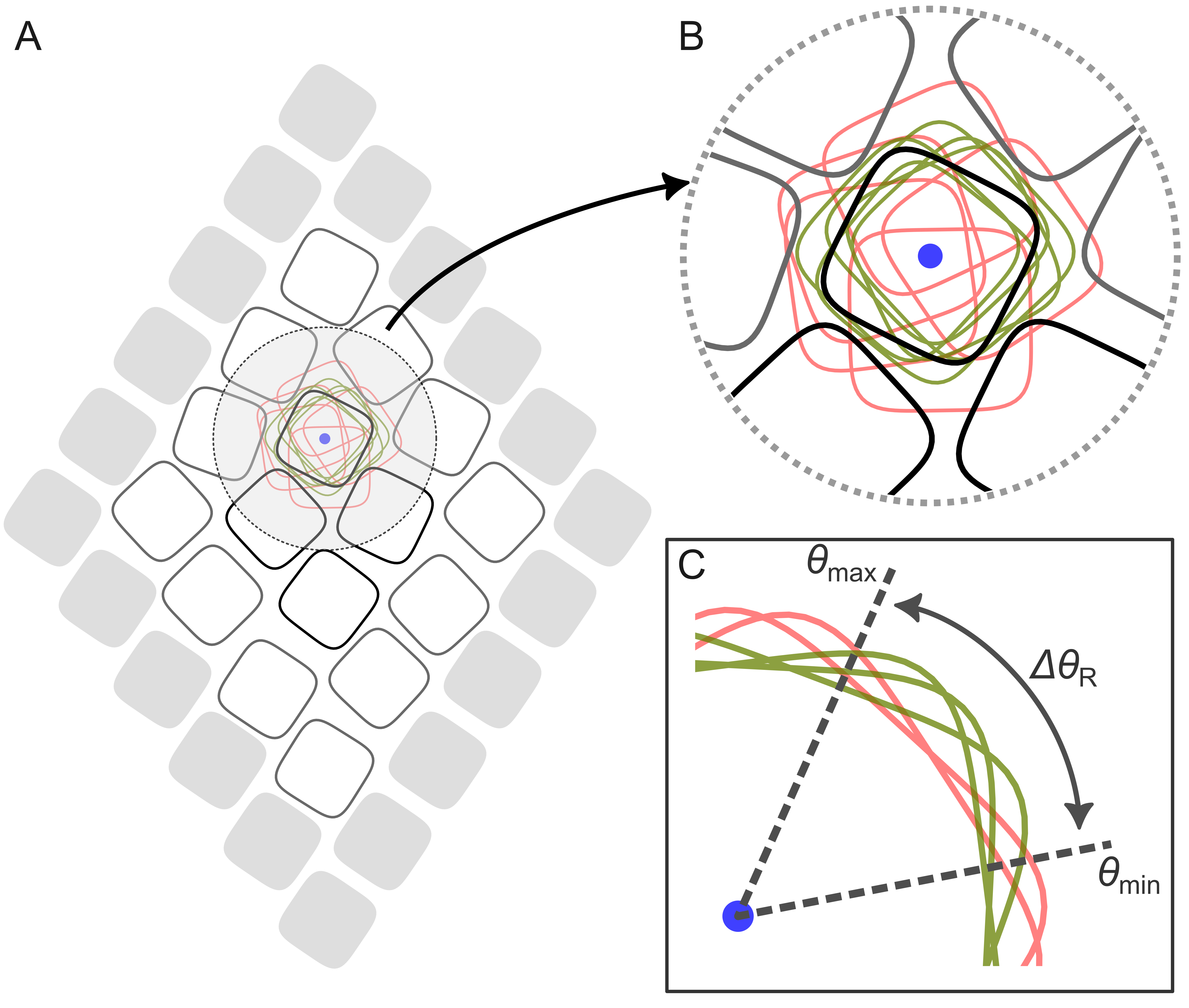}
	\caption{Accepted (green) and rejected (red) configurations of hard-interacting square particles in the BD simulations. (A) The position and orientation of each of the 16 square particles with black borders were updated at each time step. Immobile outer particles (grey filled squares) set up the bounding box, defining $\alpha$ and $\Phi$. The boundary of these square particles are defined by $|x|^p + |y|^p = (L/2)^p$, and they replicate the size and shape of the experimental particles with $L = 2.50$ length unit (\SI{1}{\micro\meter}) and $p = 4.84$. A few positional and rotational displacements of a typical particle are shown, while the green ones represent allowed configurations of the particle, and the red ones denote forbidden moves due to overlap with the neighbors. (B) The updated configurations are shown in a magnified view for clarity. (C) Updated orientations for which the range of orientational fluctuations ($\theta(t)_{\text{max}} - \theta(t)_{\text{min}}$) remained within a specified value of $\Delta \theta_{\text{R}}$ were accepted (green) and the ones that did not satisfy this criterion were rejected (red).}
	\label{BDSimulation}
\end{figure}

The initial positions and orientations of the 16 mobile particles were chosen randomly and updated after every time-step ($dt$ = \SI{2.5E-3}{\second}) following overdamped Langevin equations:
\begin{align*}
	\vec{r}_{j, t + dt} & = \vec{r}_{j, t} + \delta\vec{r}_{j, t} \\
	\theta_{j, t + dt}  & = \theta_{j, t} + \delta \theta_{j, t}
\end{align*}
where $\vec{r}_{j, t}  (= x_{j, t} \hat{x} + y_{j, t} \hat{y})$ and $\theta_{j, t}$ are the position and orientation of the $j$-th particle at time $t$; $\vec{\delta r}_{j, t} (= \delta x_{j, t} \hat{x} + \delta y_{j, t} \hat{y})$ and $\delta\theta_{j, t}$ are the Brownian translational and rotational displacements drawn from the corresponding normally distributed random series with standard deviation $\sqrt{2D_{\text{t,r}} dt}$, $D_{\text{t,r}}$ being the translational and orientational diffusion coefficients, respectively. These diffusion coeffiienets were obtained from the short time lag regions of the experimental MSDs. An updated configuration of the particles was considered valid if (i) there was no overlap between two neighboring particles ensuring excluded volume interaction and (ii) the accessible range of rotational fluctuations $(\theta(t)_{\text{max}} - \theta(t)_{\text{min}})$ of all particles remained less than a set value for the range of orientational states, $\Delta \theta_{\text{R}}$ (Fig. \ref{BDSimulation}). The values of $\Delta \theta_{\text{R}}$ for the cases of RB-C, RB-NC, HX-C, HX-NC were taken from the respective experimentally observed most-probable values (Fig. 2E and 2F). The simulations were run until the positional distributions of the particles stopped evolving with time (\textit{i.e.}, with additional simulation steps), which occurred after $\approx 7\times 10^5$ steps, corresponding to \SI{1750}{\second} in real time.

The structural ordering in the time-averaged (over 4500 steps, corresponding to $\equiv$ 11.25 s in real time) configurations upon equilibration was determined by global positional and orientational order parameters ($\Psi_4$ \& $\text{Z}_4$ for RB, and $\Psi_6$ \& $\text{Z}_6$ for HX) computed over the 4 central particles. For details regarding order parameter calculations, see \cref{sec:order}. The remaining mobile particles were excluded from the order parameter calculations because the static boundary particles partially affected their dynamics. Each simulation was repeated with six independently generated initial random configurations and the structural orderings of the time-averaged equilibrated configurations were consistent.

\begin{figure}[!htb]
	\centering
	\includegraphics[width=1\linewidth]{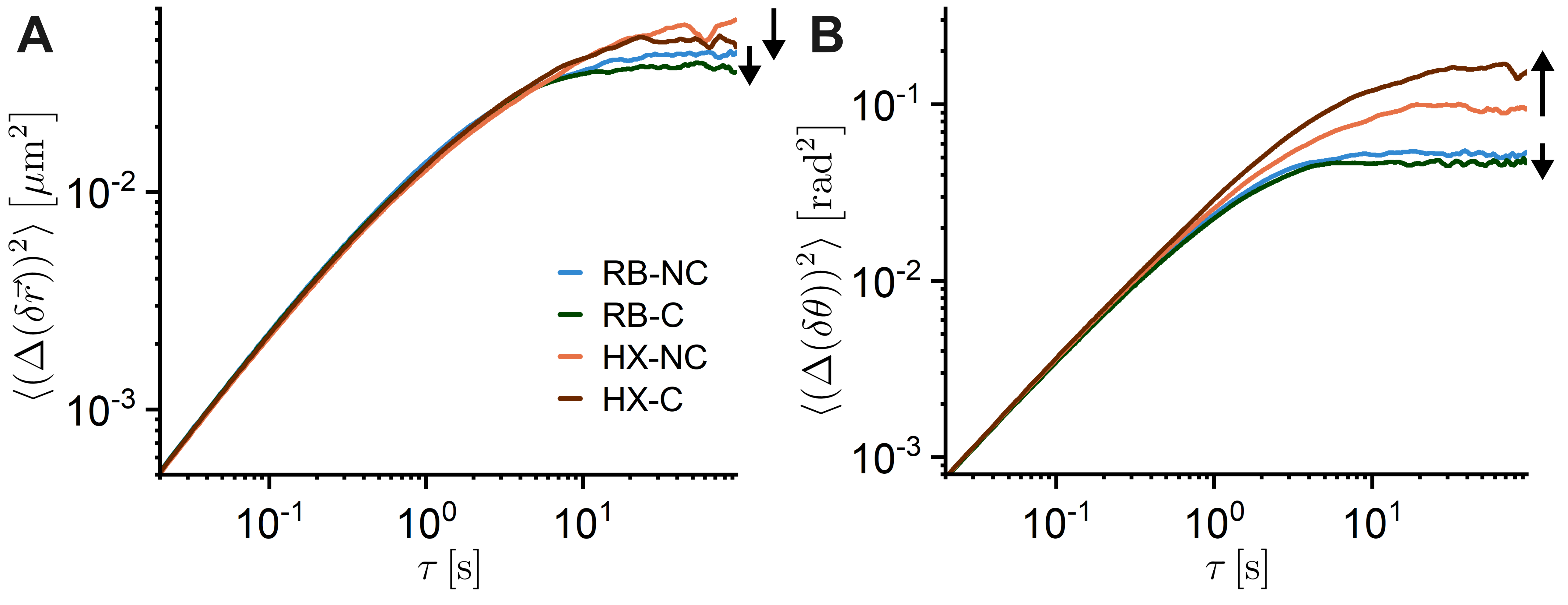}
	\caption{CR-MSDs calculated from simulated particle dynamics. (A) Translational and (B) orientational CR-MSDs for HX-NC (orange), HX-C (brown), RB-NC (blue), and RB-C (green) are shown. The black arrows indicate the changes in the corresponding CR-MSDs at the longest time lag upon crystallization. Standard errors in the CR-MSDs are negligibly small, hence are not shown.}
	\label{CRMSD_Sim}
\end{figure}

We further calculated the CR-MSD for all four simulated structural configurations, namely, RB-NC, RB-C, HX-NC, and HX-C. The translational and orientational CR-MSDs ($\left\langle (\Delta(\delta \vec{r}))^2 \right\rangle$ and $\left\langle (\Delta(\delta \theta))^2 \right\rangle$ respectively) averaged over the last $3 \times 10^5$ frames ($\equiv$ 750 s in real time) and the 4 central particles, are shown in Fig. \ref{CRMSD_Sim}. The trends of the CR-MSDs are in full agreement with those obtained from the experiments (Fig. 2 (B \& D)). The small differences in the plateau values of the CR-MSDs are attributed to a smaller system size and the complete exclusion of discrete displacements that led to overlap of neighboring particles, in the simulations.

To decouple the effect of $\Delta \theta_{\text{R}}$ from other system parameters, specifically the local density ($\Phi$) and the influence of the neighboring structural symmetry, defined by the bond angle ($\alpha$) of the bounding box, in governing the route to crystallization, we performed another BD simulation. In this case, we simulated the same system of 16 hard-interacting Brownian squares confined by 20 immobile particles, setting $\Phi = 0.77$ and $\alpha = 1.19$ rad, specifically, the same initial configuration as that for RB (Fig. 3 (B)). However, the range of accessible orientational states were relaxed as: $\Delta \theta_{\text{R}}$ = 1.02 rad, corresponding to the most-probable values of $\Delta \theta_{\text{R}}$ in HX-C. To achieve a wider $\Delta \theta_{\text{R}}$ (= 1.02 rad), which corresponds to HX-C, at a higher $\Phi$ (= 0.77), resembling RB (Fig. 3 (B)), the corner-rounding of the square particles was enhanced by reducing the value of $p$ from 4.84 to 4.50 (Fig. 4, Inset). The system was equilibrated following the same procedure and criteria as those used in previous simulations. Intriguingly, the time-averaged equilibrium configuration exhibited HX-C ordering, as identified by global bond-orientational and positional order parameters. Fig. 4 compares this equilibrium configuration with a previous simulation result where the same initial configuration equilibrated to RB-C with $\Delta \theta_{\text{R}}$ = 0.60 rad (Fig. 3 (B)).

\section{Calculation of free energy}

We studied the change in free energy as a function of the range of accessible orientational states ($\Delta \theta_{\text{R}}$) of square particles to understand how $\Delta \theta_{\text{R}}$ governs the configurational dynamics of the particles, leading them to crystallize into different preferred symmetries. The Gibbs free energy ($G$) of this 2D system can be given by $G = F + \Pi A$, where $F = U - TS$ is the Helmholtz free energy and $\Pi$ and $A$ are the 2D osmotic pressure and area, respectively. For hard-interacting ($U = 0$) particles, $F$ has only an entropic contribution, \textit{i.e.}, $F = - TS$, where $U$, $S$, and $T$ are the internal energy, entropy, and temperature, respectively. The additional term in $G$, \textit{i.e.}, $\Pi A$, becomes relevant when the change in free energy is accompanied by a variation in $\Pi$, $\Phi$, or both, such as the transition of this system of Brownian squares from a low-area fraction isotropic phase at lower osmotic pressure to a crystalline phase at a higher $\Pi$ and $\Phi$ \cite{Zhao2011}. Therefore, $\Delta G$ and $\Delta F$ exhibit different values across these transitions. However, here, we are interested in understanding the configurational dynamics, specifically the role of $\Delta \theta_{\text{R}}$ therein, which leads to the transition from HX-NC to HX-C and RB-NC to RB-C, both of which occur at the same $\Pi$ and cause an insignificant change in $\Phi$. Hence, $\Delta G (\Delta \theta_{\text{R}})$ is practically the same as $\Delta F (\Delta \theta_{\text{R}})$ for these NC-to-C transitions; therefore, $\Delta F (\Delta \theta_{\text{R}})$ was used as the relevant free energy in our study. For a system of asymmetric particles, for example, squares, the Helmholtz free energy consists of both translational ($F_{\text{t}}$) and rotational ($F_{\text{r}}$) contributions arising from the configuration entropy associated with the corresponding dynamics. The rotational part becomes irrelevant for a symmetric particle such as a disk in two dimensions.

\subsection{Variation in osmotic pressure along the cuvette length}

\begin{figure}[!th]
	\centering
	\includegraphics[width=\linewidth]{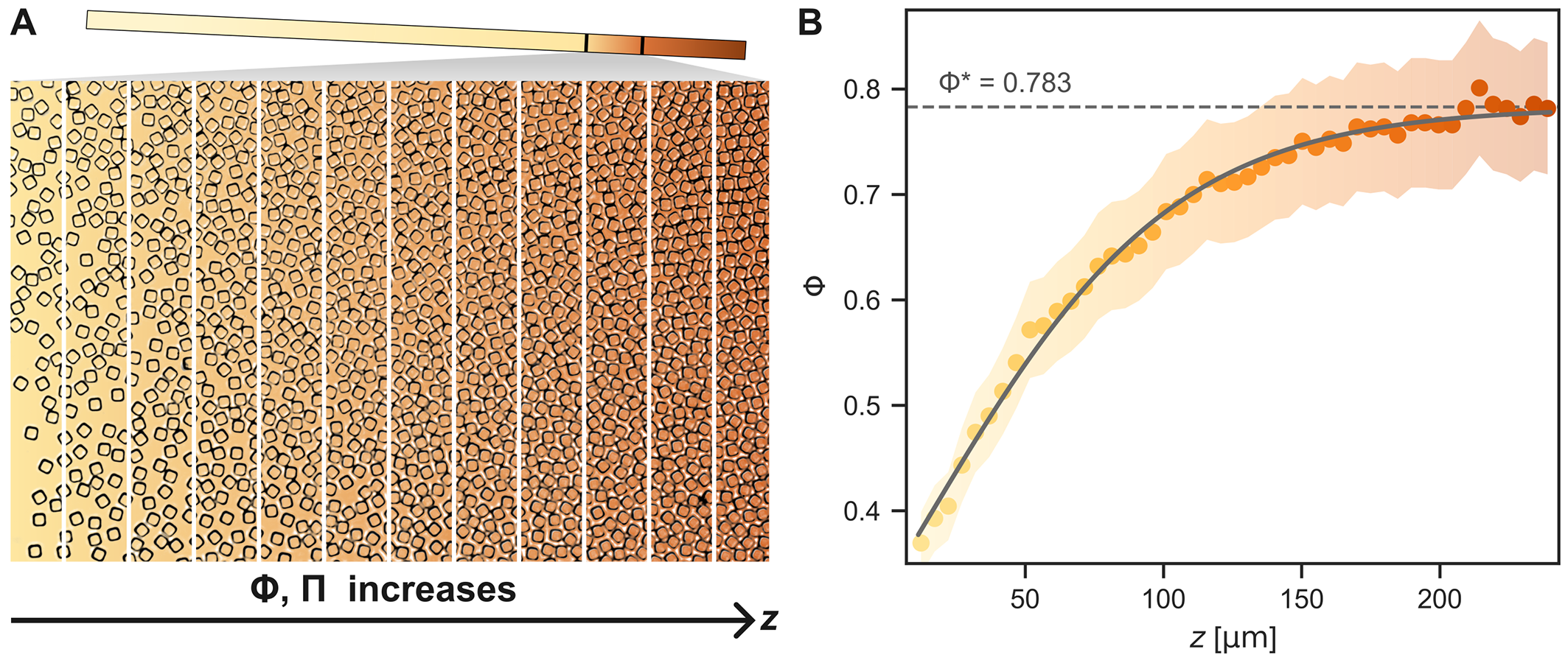}
	\caption{Variation in area fraction and osmotic pressure along the cuvette length ($z$). (A) The area fraction of particles ($\Phi$) and osmotic pressure ($\Pi$) increase along the length ($z$) of an inclined cuvette (not in scale), from top to bottom. The number of strips into which the image frame is divided is equal to the number of data points in the plot shown in (B); only a few strips are shown here for illustration. (B) The plot of $\Phi(z)$ is shown with filled circles of varying colors, indicating their position on the image frame and in the cuvette, as shown in (A). Estimated errors in deriving $\Phi(z)$ from captured image frames are depicted by the shaded region. The solid line is the best fit to the plot following Eq. \ref{eq:phi} and the dashed horizontal line indicates $\Phi^{\ast}$, which is the height of the eventual plateau of $\Phi(z)$.}
	\label{OsmoticPressure}
\end{figure}

Because our system was under osmotic compression on an inclined plane, the number density of particles ($\rho$), and hence, the area fraction ($\Phi$), increased monotonically with $\Pi$ towards the bottom (Fig. \ref{OsmoticPressure}). Their dependency can be expressed analytically as \cite{Wang2015}:
\begin{equation}
	\Phi(z) = \Phi^{\ast} \left[ 1 + \exp(-(z-z_0)/h) \right]^{-1} \label{eq:phi},
\end{equation}
where $\Phi^{\ast}$ is the value of $\Phi$ after it reaches a plateau, corresponding to the densest packing at the bottom end of the cuvette; $z$, $z_0$, and $h$ denote the distance along the cuvette from top to bottom, a reference position, and the gravitational height of the system, respectively.

We captured an image frame with a wider field of view using a lower-magnification ($10 \times$) objective and a DSLR camera with a larger sensor to observe the variation in $\Phi$ along $z$ for our system. By dividing the image frame into thin slices at varying $z$-values and determining the local area fraction of each strip, we obtained adequate data points for plotting $\Phi(z)$. The $\Phi(z)$ plot and the fitted exponential growth following Eq. \ref{eq:phi} are shown in Fig. \ref{OsmoticPressure}. This provides the value of the fitting parameter $\Phi^{\ast}$ as $\Phi^{\ast} = 0.783$.

We obtained the osmotic pressure at distance $z$ by summing the contributions from all the particles above $z$ \cite{Wang2015}. Thus,

\begin{align}
	\Pi(z) & = \frac{k_{\text{B}}T}{A_{\text{p}}} \frac{1}{h} \int_{-\infty}^{z} \Phi(z')dz' \label{eq:piz}                            \\
	& = \frac{k_{\text{B}}T}{A_{\text{p}}} \frac{1}{h} \int_{-\infty}^{z} \Phi^{\ast} \left[ 1 + \exp(-(z'-z_0)/h) \right]^{-1} dz' \nonumber\\
	& = \frac{k_{\text{B}}T}{A_{\text{p}}} \times \Phi^{\ast} \ln \left[ 1 + \exp((z-z_0)/h) \right] \label{eq:pi}
\end{align}
where $k_{\text{B}}$ and $A_{\text{p}}$ are the Boltzmann constant and area of each particle, respectively.

Using Eq. \ref{eq:phi}, the parametric relation between $\Pi$ and $\Phi$ can be expressed as 

\begin{align}
	\tilde{\Pi}(\Phi) & = \frac{\Pi(\Phi)}{(k_{\text{B}}T/A_{\text{p}})} =  \Phi^{\ast} \ln \left( \frac{\Phi^{\ast}}{\Phi^{\ast} - \Phi} \right).  \label{eq:piphi}
\end{align}

This relation provides the values of $\tilde{\Pi}_{\text{RB}}$ and $\tilde{\Pi}_{\text{HX}}$ for the two phases RB and HX using their area fraction values $\Phi_{\text{RB}} = 0.782  \pm 0.001$ and $\Phi_{\text{HX}} = 0.768  \pm 0.001$, respectively. Thus, using $\Phi^{\ast} = 0.783$, we obtain $\tilde{\Pi}_{\text{RB}} = 5.77$ and $\tilde{\Pi}_{\text{HX}} = 3.12$.

\subsection{Variation of translational free energy with the range of accessible orientational states}

The translational part of the free energy $F_{\text{t}}$  was calculated using thermodynamic integration. The 2D osmotic pressure $\Pi$ can be expressed as a partial derivative of $F_{\text{t}}$ with respect to area ($A$) at a constant particle number ($N$), $ \Pi = - \frac{\partial F_{\text{t}}}{\partial A}|_N$ \cite{Zhao2011}. Therefore, translational free energy ($F_{\text{t}}$) can be obtained using the following integration:

\begin{align}
	\Delta F_{\text{t}} & = N \int \frac{\Pi(\rho')}{\rho'^2} d\rho'         \nonumber                        \\
	& = N A_{\text{p}} \int \frac{\Pi(\Phi')}{\Phi'^2} d\Phi'  
	\label{eq:ftin}
\end{align}
where the number density ($\rho$) and area fraction ($\Phi$) are defined as $\rho = N/A$ and $\Phi = N A_{\text{p}} / A = A_{\text{p}} \rho$, respectively; $\rho'$ and $\Phi'$ are the corresponding integration variables.

The osmotic pressure $\Pi$ is related to the area fraction $\Phi$ by Eq. \ref{eq:piphi}. However, when using Eq. \ref{eq:piphi} to replace the integrand in Eq. \ref{eq:ftin}, $\Phi^{\ast}$ must be replaced by $\Phi_{\text{max}}$, which is the maximum possible area fraction that the square particles can attain for a given value of $\Delta\theta_{\text{R}}$. It is important to note that $\Phi_{\text{max}}(\Delta\theta_{\text{R}})$ represents the theoretical maximum value of $\Phi$, corresponding to a reference state in which the particles have adequate accessible free space to allow the rotational fluctuations of the range $\Delta \theta_{\text{R}}$ and nonexistent translational movements. Moreover, $\Phi_{\mathrm{max}}$ depends strongly on $\Delta \theta_{\text{R}}$ because of interdigitation, and hence, is governed by the degree of corner rounding of the square particles. Following this and Eq. \ref{eq:piphi}, we obtain,

\begin{align}                                           
	& F_{\text{t}} = N A_{\text{p}} \frac{k_{\text{B}}T}{A_{\text{p}}} \Phi_{\mathrm{max}}(\Delta \theta_{\text{R}}) \int \frac{1}{\Phi'^2} \ln \left( \frac{\Phi_{\mathrm{max}}(\Delta \theta_{\text{R}})}{\Phi_{\mathrm{max}}(\Delta \theta_{\text{R}}) -\Phi'} \right) d\Phi' \tag{from Eq. \ref{eq:ftin}}   \nonumber       \\
	& \Rightarrow \frac{F_{\text{t}}}{Nk_{\text{B}}T} = \Phi_{\mathrm{max}}(\Delta \theta_{\text{R}}) \int \frac{1}{\Phi'^2} \ln \left( \frac{\Phi_{\mathrm{max}}(\Delta \theta_{\text{R}})}{\Phi_{\mathrm{max}}(\Delta \theta_{\text{R}}) -\Phi'} \right) d\Phi'              \nonumber                               \\
	& \Rightarrow \frac{\Delta F_{\text{t}}}{Nk_{\text{B}}T} = \ln \left( \frac{\Phi}{\Phi_{\mathrm{max}}(\Delta \theta_{\text{R}}) - \Phi} \right) - \frac{\Phi_{\mathrm{max}}(\Delta \theta_{\text{R}})}{\Phi} \ln \left( \frac{\Phi_{\mathrm{max}}(\Delta \theta_{\text{R}})}{\Phi_{\mathrm{max}}(\Delta \theta_{\text{R}}) - \Phi} \right)  \nonumber \\
	& \Rightarrow \frac{\Delta F_{\text{t}}}{Nk_{\text{B}}T} = \ln \left[ \exp(\tilde{\Pi}\Phi_{\mathrm{max}}^{-1}(\Delta \theta_{\text{R}})) - 1 \right] - \tilde{\Pi}\Phi^{-1}  \label{eq:ft}
\end{align}

This Eq. \ref{eq:ft} represents the change in the translational free energy in this system, and the second term, which is equal to $\Pi A$, is the difference between $\Delta F$ and $\Delta G$, as discussed above. For transitions in which $\Delta (\Pi A)$ is finite, this term becomes part of the Gibbs free energy ($G = F + \Pi A$). However, in this study of NC-to-C transitions, $\Delta (\Pi A) \simeq 0$, and the change in translational free energy is given only by the first term on the RHS of Eq. \ref{eq:ft}.

The change in rotational free energy arises from the different orientational states that a particle can access for a given orientation range, and is expressed as,
\begin{equation}
	\Delta F_\text{r}/Nk_{\text{B}}T = -\ln(4\Delta\theta_{\text{R}}).
	\label{eq:fr}
\end{equation}

Therefore, we obtain the variation in the dimensionless free energy per particle, $\Delta F/Nk_{\text{B}}T$, with $\Delta\theta_{\text{R}}$ as:

\begin{align}
	& \Delta F/Nk_{\text{B}}T = \Delta F_{\text{t}}/Nk_{\text{B}}T + \Delta F_{\text{r}}/Nk_{\text{B}}T     \nonumber                                                                                                                                                 \\
	& \Rightarrow \frac{\Delta F}{Nk_{\text{B}}T} = \ln \left[ \exp(\tilde{\Pi}/\Phi_{\text{max}}(\Delta\theta_{\text{R}})) - 1 \right] - \ln(4\Delta\theta_{\text{R}})  \label{eq:df}
\end{align}

\subsection{Simulating the dependence of $\Phi_{\mathrm{max}}$ on the range of accessible orientational states $(\Delta\theta_{\mathrm{R}})$}

As discussed above, $\Phi_{\text{max}}(\Delta\theta_{\text{R}})$ is the theoretical maximum value of $\Phi$, which allows orientational fluctuations in the range of $\Delta\theta_{\text{R}}$. The dependence $\Phi_{\text{max}}(\Delta\theta_{\text{R}})$ was derived from a simulation. We considered four corner-rounded square particles (corresponding to $p = 4.84$, resembling the experimental particles, as shown in Fig. 4, inset) and packed them as densely as possible allowing a given range of orientational fluctuations.

\begin{figure}[!htb]
	\centering
	\includegraphics[width=0.75\linewidth]{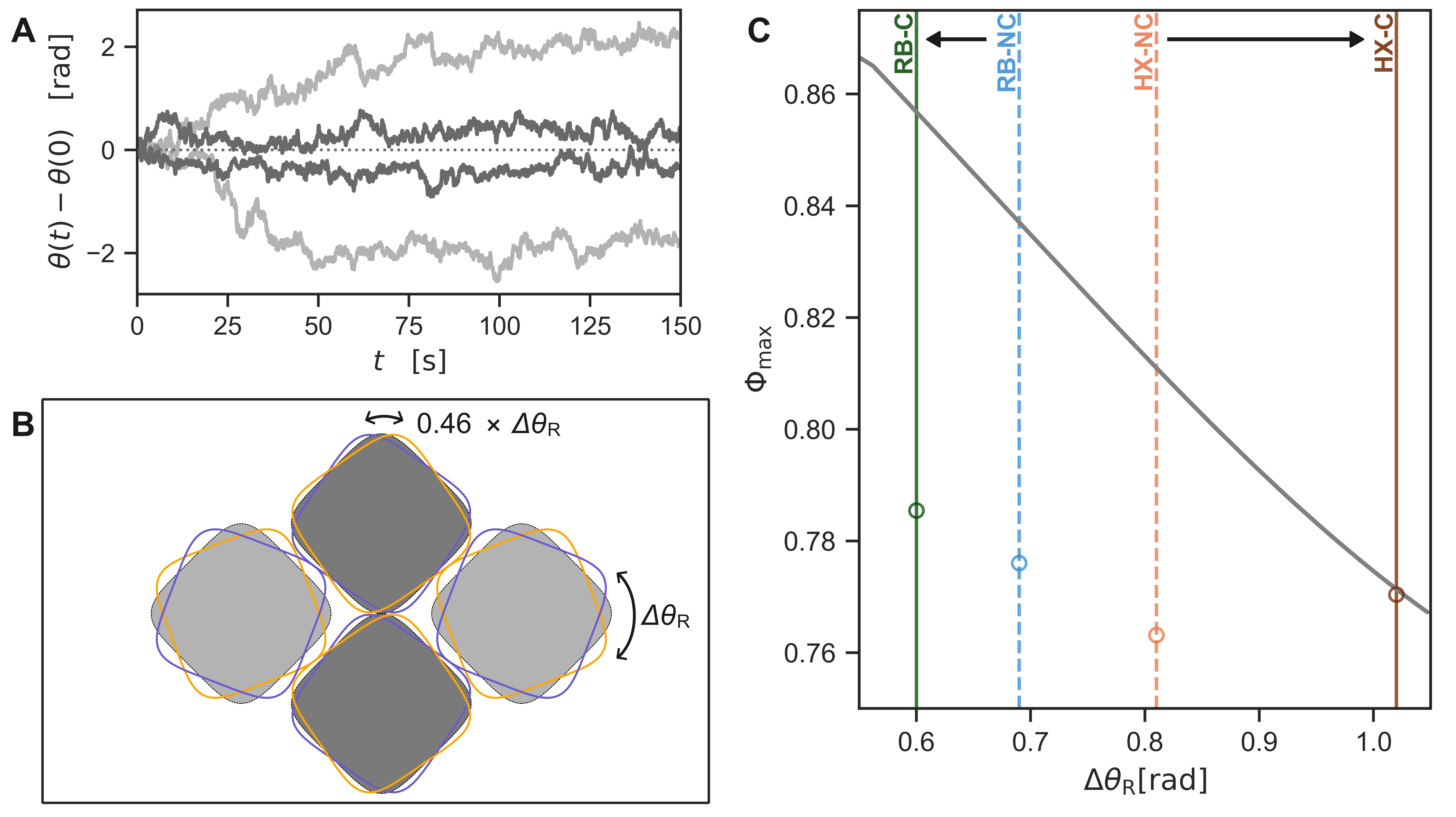}
	\caption{Range of accessible orientational states of neighboring particles and the dependence of $\Phi_{\text{max}}$ on $\Delta\theta_{\text{R}}$. (A) Orientational trajectories, \textit{i.e.}, instantaneous orientations, $\theta(t)$, in reference to the initial orientations, $\theta(0)$, of the four particles are plotted against time, $t$. (B) Schematic shows the configuration of the four particles and the range of their accessible rotational states. While the pair of diagonal particles that have a wider range ($\Delta\theta_{\text{R}}$) is shown in light grey, the pair with a narrower range ($0.46 \times \Delta\theta_{\text{R}}$) is displayed in dark grey, in accordance with the color of their respective orientational trajectories. The blue and orange squares denote the extreme accessible orientations of the particles. (C) Variation in $\Phi_{\text{max}}$ with $\Delta\theta_{\text{R}}$ is shown. The solid and dashed vertical lines denote the $\Delta\theta_{\text{R}}$ values corresponding to the C and NC regions, respectively, in both the RB and HX. The colors of the lines are as per the consistent color scheme used in other figures to represent the four different structural phases, and the black arrows show the change in $\Delta\theta_{\text{R}}$ upon crystallization in RB and HX. The open circles on the vertical lines mark the respective values of the area fraction ($\Phi$) observed in the experiments.}
	\label{AFmax}
\end{figure}

In the experimental data, the orientational time-series of neighboring squares at higher area fractions (RB or HX) were observed to have reciprocal ranges,\textit{ i.e.}, if a particle had a wider range of accessible rotational states, its neighbors would have narrower orientational fluctuations at that time. This process is illustrated in Fig. \ref{AFmax} by plotting the orientational trajectories of 4 neighboring particles in pane (A) and showing the configuration of the corresponding particles (dark and light grey) in pane (B). Therefore, not all particles can simultaneously access the complete range of orientational states given by $\Delta\theta_{\text{R}}$. We analyzed the average behavior of the simultaneous orientational trajectories of neighboring particles and deduced that while some of the particles accessed a wider range of orientational fluctuations $\Delta\theta_{\text{R}}$, the immediate neighbors of those particles had access to a narrower range of $\lesssim 0.5 \times \Delta\theta_{\text{R}}$ at that time.

Following this experimentally observed trend, we restricted the orientational fluctuations of the two diagonal particles within $\Delta\theta_{\text{R}}$ and those of the other two particles to the range of $0.46 \times \Delta\theta_{\text{R}}$, adhering to the configuration shown in Fig. \ref{AFmax} (B). We chose to use $0.46 \times \Delta\theta_{\text{R}}$ as the reduced range for the neighboring particles because it is closest to the observation ($\lesssim 0.5 \times \Delta\theta_{\text{R}}$), providing a feasible value of $\Phi_{\text{max}}$ at $\Delta\theta_{\text{R}}$ corresponding to HX-C (Fig. \ref{AFmax} (C)).

In the initial configuration, four corner-rounded square particles were placed at the vertices of a rhombus, avoiding overlap and aligning their diagonals with those of the rhombus. This configuration was chosen because it led to the densest packing of square particles. Starting from this arrangement, we brought the particles closer to each other with small inward displacements in each step until they overlapped while accessing their set range of rotational fluctuations. We checked for possible overlap by rotating each particle over the orientational range (${- \Delta\theta_{\text{R}}}n/2$, ${\Delta\theta_{\text{R}}}n/2$), where the multiplicative factor $n = 1$ for a pair of diagonal particles and $n = 0.46$ for the other pair of particles (Fig. \ref{AFmax} (B)). The area fraction at the densest packing without overlap was recorded as the $\Phi_{\text{max}}$ value for this $\Delta\theta_{\text{R}}$. The simulation was performed at various closely spaced values of $\Delta\theta_{\text{R}}$ to obtain the dependence of $\Phi_{\text{max}}(\Delta\theta_{\text{R}})$ (Fig. \ref{AFmax} (C)).

\subsection{Variation in free energy with $\Delta\theta_{\mathrm{R}}$ in the CE phase}

\begin{figure}[!hbt]
	\centering
	\includegraphics[width=0.6\linewidth]{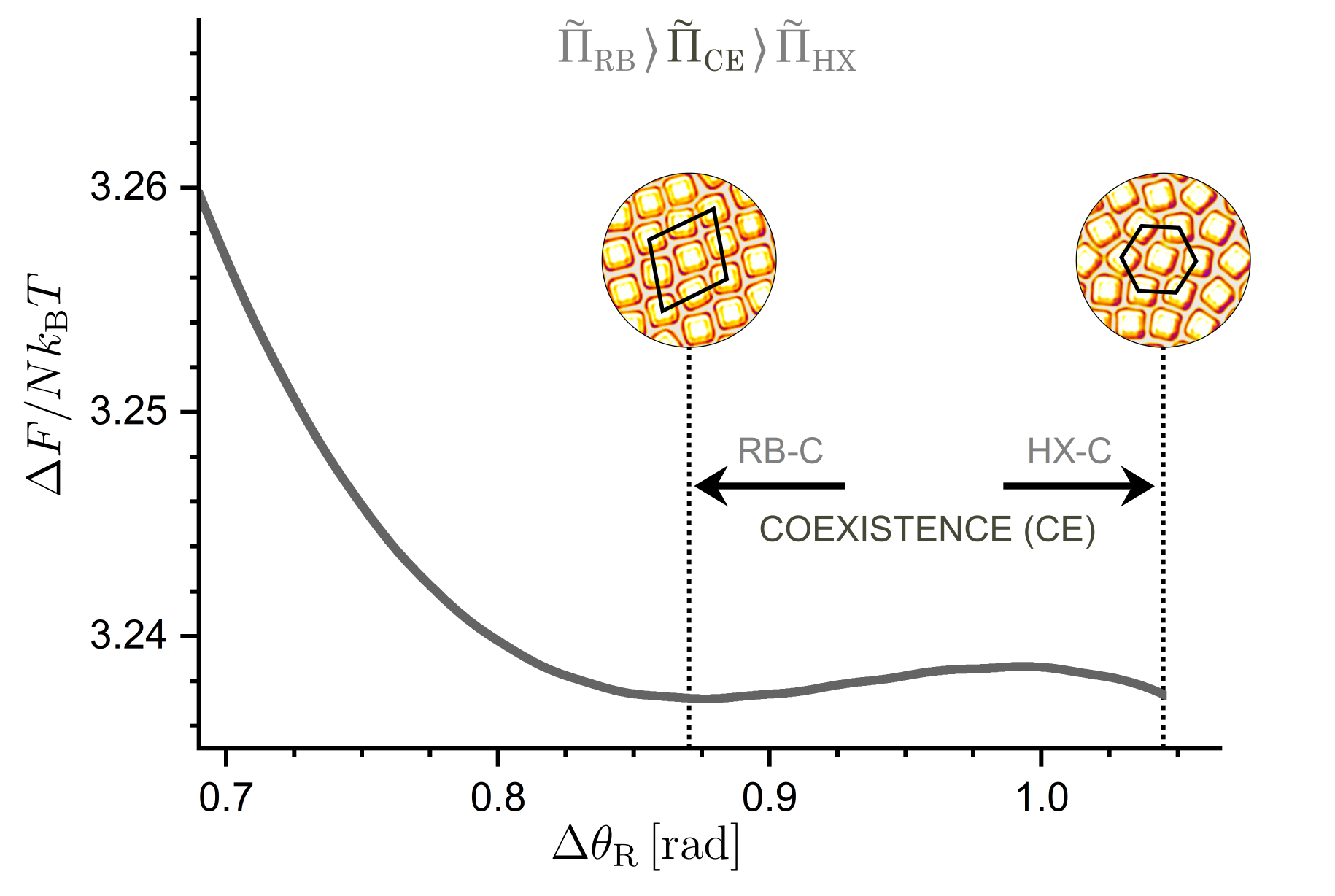}
	\caption{Free energy variation with $\Delta\theta_{\text{R}}$ in the CE phase. The variation in the dimensionless free energy per particle ($\Delta F/Nk_{\text{B}}T$) with the range of accessible orientational states ($\Delta\theta_{\text{R}}$) is shown for $\tilde{\Pi}_{\mathrm{CE}} = 3.59$, which is intermediate between $\tilde{\Pi}_{\text{RB}}$ and $\tilde{\Pi}_{\text{HX}}$, representing the CE phase. The dashed vertical lines mark the two minima that emerge at the $\Delta\theta_{\text{R}}$ values corresponding to RB-C and HX-C. Their respective structural orderings are shown as circular insets on the top of the vertical lines.}
	\label{FE_CE}
\end{figure}

The RB and HX phases studied here are formed at the two boundaries of the coexistence phase (CE); therefore, CE is observed at osmotic pressure values intermediate to $\tilde{\Pi}_{\text{RB}}$ and $\tilde{\Pi}_{\text{HX}}$. We computed the variation in the dimensionless free energy per particle ($\Delta F/Nk_{\text{B}}T$) with the range of accessible orientational states of the particles, $\Delta\theta_{\text{R}}$, at $\tilde{\Pi}_{\text{CE}} = 3.59$, which is between $\tilde{\Pi}_{\text{RB}} = 5.77$ and $\tilde{\Pi}_{\text{HX}} = 3.12$. This variation in $\Delta F/Nk_{\text{B}}T$ (Fig. \ref{FE_CE}) shows a broad peak with two minima on the two sides at $\Delta\theta_{\text{R}}$ values corresponding to RB-C and HX-C. The broad peak in the free energy variation explains the existence of the CE phase, where a local perturbation in orientational dynamics,\textit{ i.e.}, in $\Delta\theta_{\text{R}}$, induces either RB or HX ordering as the local region is pushed to the left or right minima, respectively.

\subsection{Applicability to Similar Systems}

We also simulated the dependence of $\Phi_{\textrm{max}}$ on $\Delta\theta_\textrm{R}$ for another similar system, \textit{i.e.}, triangular platelet particles \cite{Zhao2012a}, and a similar monotonic decrease in $\Phi_{\textrm{max}} (\Delta\theta_\textrm{R})$, as observed for squares, was obtained. Both of these variations are shown in Fig. 7. The simulated structural configurations and their ordering at the densest area fraction for various values of $\Delta\theta_\textrm{R}$ completely agree with the experimental observations in both the cases (Fig. 7, insets).

\section{Effect of hydrodynamic forces on crystallization pathways}
In our experimental system, the square particles (length $L = 2.5$ \unit{\micro\meter} and thickness $d = 1$ \unit{\micro \meter}) form a monolayer close to the bottom surface of the cuvette, where hydrodynamic drag on the lithographic particles is dominated by the lubrication interaction between the cuvette wall and particle face nearer to it; which in-turn effectively suppresses any in-plane hydrodynamic forces between particles. This is primarily because of two factors: 
(i) the separation between the particle bottom face and the wall is significantly smaller ($h < 40$ nm) than the inter-particle separation ($\epsilon \gtrsim 300$ nm), and 
(ii) the face surface area is larger than those of the edges; and can be demonstrated using the grand resistance matrix in the lubrication approximation \cite{Swan2007}, $\mathcal{R} \approx \mathcal{R}^{\mathrm{pw}} + \mathcal{R}^{\mathrm{pp}}$, where $\mathcal{R}^{\mathrm{pw}} = \begin{bsmallmatrix}
	R^{\mathrm{pw}} & 0\\
	0 & R^{\mathrm{pw}}
\end{bsmallmatrix}$ and $\mathcal{R}^{\mathrm{pp}} = \begin{bsmallmatrix}
	R^{\mathrm{pp}} & -R^{\mathrm{pp}}\\
	-R^{\mathrm{pp}} & R^{\mathrm{pp}}
\end{bsmallmatrix}$ denote the resistance matrices arising from particle–wall and particle–particle lubrication forces, respectively. 
So, the self- and cross-mobility can be given as, $\mu_{\mathrm{self}} =\frac{R^{\mathrm{pw}}+R^{\mathrm{pp}}}{R^{\mathrm{pw}} (R^{\mathrm{pw}}+ 2R^{\mathrm{pp}})}$ and $\mu_{\mathrm{cross}} =\frac{R^{\mathrm{pp}}}{R^{\mathrm{pw}} (R^{\mathrm{pw}}+ 2R^{\mathrm{pp}})}$.
For a model system of two square platelets near a wall, where the particles are perfectly side-aligned as they approach each other, the resistance due to viscous drag (shear) is given by $R^{\mathrm{pw}} = \eta A_1/h$, and we approximate the resistance due to two approaching square particles as $R^{\mathrm{pp}} = 3\eta A_2^2/2\pi\epsilon^3$ (derived by appropriately modifying the expression for two approaching disks \cite{Nguyen2000}), where $A_1=L^2$, $A_2=Ld$ are the surface areas corresponding to the face and edge of the square particles.

In our case, using $h = 10$ \unit{\nano\meter}, the self-diffusivity, $D = \mu_{\mathrm{self}} k_{\mathrm{B}} T = \frac{h k_{\mathrm{B}} T}{\eta L^2 }\cdot \frac{3hd^2/2\pi \epsilon^3 + 1}{3hd^2/\pi \epsilon^3 + 1}$ of square particles turns out to be  $0.0058$ \unit{\square\micro\meter\per\second}. This matches well with the experimental diffusion coefficient, $D_{\mathrm{t}} = 0.0063$ \unit{\square\micro\meter\per\second}, obtained from the MSD of the particles. \textit{Nevertheless, reduced mobilities of the square particles only make their dynamics slower but do not affect their crystallization pathways.} The crystallization pathways remain solely dependent on the free energy landscape, and for our entropic colloidal system, it is determined by the configurational entropy of the system.

Similarly, the ratio of cross- to self-mobility, which determines the relative contribution of in-plane particle-particle hydrodynamic interactions towards the motion of a particle, is obtained to be $R^{\mathrm{pp}}/(R^{\mathrm{pp}}+R^{\mathrm{pw}})=\left(1 + 2\pi \epsilon ^3/3hd^2\right)^{-1} \approx 15 \%$. 
Additionally, we must consider other factors such as the roughness of the particle surfaces and the relative orientation of the particles. The fluid in the thin layer between the edges can escape more easily through channels formed by asperities, resulting in less pressure buildup and consequently a decrease in the normal resistance $R^{\mathrm{pp}}$. In contrast, the tangential resistance ($R^{\mathrm{pw}}$) increases as the area in contact with the wall is greater when the particle face is rough. So, the effective cross- to self-mobility ratio gets reduced further. Finally, it is important to note that the model system considered above is an extreme case in which the edges of square platelets are completely parallel. In reality, particles in the RB-NC phase are rarely aligned, resulting in a sharp decrease in $R^{\mathrm{pp}}$; similarly, in the HX-NC, HX-C domains, the separation of particles is comparatively greater, and their side-wise alignment is even less frequent. Considering the combined effects of the aforementioned points, we have neglected any in-plane hydrodynamic contributions in the crystallization dynamics, i.e., from RB-NC to RB-C and from HX-NC to HX-C. Although there could be shape- and orientation-dependent in-plane drag contributions that regulate both the translational and rotational dynamics of the particle, these will be negligibly small compared to the facial hydrodynamic interactions. \textit{We again emphasize that even negligible in-plane drag contributions can slightly affect the timescale of the dynamics and hence that of the crystallization process, but not the kinetic pathways of crystallization.}

\bibliographystyle{elsarticle-num} 
\bibliography{RouteToCrystallization_2.bib}
